\documentclass[%
 reprint,
superscriptaddress,
 amsmath,amssymb,
 aps,
]{revtex4-2}
\pdfoutput=1
\usepackage[utf8]{inputenc}
\usepackage{graphicx}
\usepackage{dcolumn}
\usepackage{bm}
\usepackage{epstopdf}
\usepackage{array}
\usepackage{makecell}
\usepackage[mathlines]{lineno}
\usepackage{xcolor}
\usepackage{booktabs}
\usepackage{upgreek}
\usepackage{appendix}
\usepackage{multirow}
\usepackage{amsmath}    
\begin{document}

\title{New Methods and Simulations for Cosmogenic Induced Spallation Removal \\ in Super-Kamiokande-IV}

\newcommand{\AFFicrr}{\affiliation{Kamioka Observatory, Institute for Cosmic Ray Research, University of Tokyo, Kamioka, Gifu 506-1205, Japan}}
\newcommand{\AFFkashiwa}{\affiliation{Research Center for Cosmic Neutrinos, Institute for Cosmic Ray Research, University of Tokyo, Kashiwa, Chiba 277-8582, Japan}}
\newcommand{\AFFicrronly}{\affiliation{Institute for Cosmic Ray Research, University of Tokyo, Kashiwa, Chiba 277-8582, Japan}}
\newcommand{\AFFipmu}{\affiliation{Kavli Institute for the Physics and
Mathematics of the Universe (WPI), The University of Tokyo Institutes for Advanced Study,
University of Tokyo, Kashiwa, Chiba 277-8583, Japan }}
\newcommand{\AFFmad}{\affiliation{Department of Theoretical Physics, University Autonoma Madrid, 28049 Madrid, Spain}}
\newcommand{\AFFubc}{\affiliation{Department of Physics and Astronomy, University of British Columbia, Vancouver, BC, V6T1Z4, Canada}}
\newcommand{\AFFbu}{\affiliation{Department of Physics, Boston University, Boston, MA 02215, USA}}
\newcommand{\AFFuci}{\affiliation{Department of Physics and Astronomy, University of California, Irvine, Irvine, CA 92697-4575, USA }}
\newcommand{\AFFcsu}{\affiliation{Department of Physics, California State University, Dominguez Hills, Carson, CA 90747, USA}}
\newcommand{\AFFcnm}{\affiliation{Institute for Universe and Elementary Particles, Chonnam National University, Gwangju 61186, Korea}}
\newcommand{\AFFduke}{\affiliation{Department of Physics, Duke University, Durham NC 27708, USA}}
\newcommand{\AFFfukuoka}{\affiliation{Junior College, Fukuoka Institute of Technology, Fukuoka, Fukuoka 811-0295, Japan}}
\newcommand{\AFFgifu}{\affiliation{Department of Physics, Gifu University, Gifu, Gifu 501-1193, Japan}}
\newcommand{\AFFgist}{\affiliation{GIST College, Gwangju Institute of Science and Technology, Gwangju 500-712, Korea}}
\newcommand{\AFFuh}{\affiliation{Department of Physics and Astronomy, University of Hawaii, Honolulu, HI 96822, USA}}
\newcommand{\AFFicl}{\affiliation{Department of Physics, Imperial College London , London, SW7 2AZ, United Kingdom }}
\newcommand{\AFFkek}{\affiliation{High Energy Accelerator Research Organization (KEK), Tsukuba, Ibaraki 305-0801, Japan }}
\newcommand{\AFFkobe}{\affiliation{Department of Physics, Kobe University, Kobe, Hyogo 657-8501, Japan}}
\newcommand{\AFFkyoto}{\affiliation{Department of Physics, Kyoto University, Kyoto, Kyoto 606-8502, Japan}}
\newcommand{\AFFliv}{\affiliation{Department of Physics, University of Liverpool, Liverpool, L69 7ZE, United Kingdom}}
\newcommand{\AFFmiyagi}{\affiliation{Department of Physics, Miyagi University of Education, Sendai, Miyagi 980-0845, Japan}}
\newcommand{\AFFnagoya}{\affiliation{Institute for Space-Earth Environmental Research, Nagoya University, Nagoya, Aichi 464-8602, Japan}}
\newcommand{\AFFkmi}{\affiliation{Kobayashi-Maskawa Institute for the Origin of Particles and the Universe, Nagoya University, Nagoya, Aichi 464-8602, Japan}}
\newcommand{\AFFpol}{\affiliation{National Centre For Nuclear Research, 02-093 Warsaw, Poland}}
\newcommand{\AFFsuny}{\affiliation{Department of Physics and Astronomy, State University of New York at Stony Brook, NY 11794-3800, USA}}
\newcommand{\AFFokayama}{\affiliation{Department of Physics, Okayama University, Okayama, Okayama 700-8530, Japan }}
\newcommand{\AFFosaka}{\affiliation{Department of Physics, Osaka University, Toyonaka, Osaka 560-0043, Japan}}
\newcommand{\AFFox}{\affiliation{Department of Physics, Oxford University, Oxford, OX1 3PU, United Kingdom}}
\newcommand{\AFFqmul}{\affiliation{School of Physics and Astronomy, Queen Mary University of London, London, E1 4NS, United Kingdom}}
\newcommand{\AFFregina}{\affiliation{Department of Physics, University of Regina, 3737 Wascana Parkway, Regina, SK, S4SOA2, Canada}}
\newcommand{\AFFseoul}{\affiliation{Department of Physics, Seoul National University, Seoul 151-742, Korea}}
\newcommand{\AFFsheff}{\affiliation{Department of Physics and Astronomy, University of Sheffield, S3 7RH, Sheffield, United Kingdom}}
\newcommand{\AFFshizuokasc}{\affiliation{Department of Informatics in
Social Welfare, Shizuoka University of Welfare, Yaizu, Shizuoka, 425-8611, Japan}}
\newcommand{\AFFstfc}{\affiliation{STFC, Rutherford Appleton Laboratory, Harwell Oxford, and Daresbury Laboratory, Warrington, OX11 0QX, United Kingdom}}
\newcommand{\AFFskk}{\affiliation{Department of Physics, Sungkyunkwan University, Suwon 440-746, Korea}}
\newcommand{\AFFtokyo}{\affiliation{The University of Tokyo, Bunkyo, Tokyo 113-0033, Japan }}
\newcommand{\AFFtodai}{\affiliation{Department of Physics, University of Tokyo, Bunkyo, Tokyo 113-0033, Japan }}
\newcommand{\AFFtit}{\affiliation{Department of Physics,Tokyo Institute of Technology, Meguro, Tokyo 152-8551, Japan }}
\newcommand{\AFFtus}{\affiliation{Department of Physics, Faculty of Science and Technology, Tokyo University of Science, Noda, Chiba 278-8510, Japan }}
\newcommand{\AFFtoronto}{\affiliation{Department of Physics, University of Toronto, ON, M5S 1A7, Canada }}
\newcommand{\AFFtriumf}{\affiliation{TRIUMF, 4004 Wesbrook Mall, Vancouver, BC, V6T2A3, Canada }}
\newcommand{\AFFtokai}{\affiliation{Department of Physics, Tokai University, Hiratsuka, Kanagawa 259-1292, Japan}}
\newcommand{\AFFtsinghua}{\affiliation{Department of Engineering Physics, Tsinghua University, Beijing, 100084, China}}
\newcommand{\AFFynu}{\affiliation{Department of Physics, Yokohama National University, Yokohama, Kanagawa, 240-8501, Japan}}
\newcommand{\AFFllr}{\affiliation{Ecole Polytechnique, IN2P3-CNRS, Laboratoire Leprince-Ringuet, F-91120 Palaiseau, France }}
\newcommand{\AFFbari}{\affiliation{ Dipartimento Interuniversitario di Fisica, INFN Sezione di Bari and Universit\`a e Politecnico di Bari, I-70125, Bari, Italy}}
\newcommand{\AFFnapoli}{\affiliation{Dipartimento di Fisica, INFN Sezione di Napoli and Universit\`a di Napoli, I-80126, Napoli, Italy}}
\newcommand{\AFFroma}{\affiliation{INFN Sezione di Roma and Universit\`a di Roma ``La Sapienza'', I-00185, Roma, Italy}}
\newcommand{\AFFpadova}{\affiliation{Dipartimento di Fisica, INFN Sezione di Padova and Universit\`a di Padova, I-35131, Padova, Italy}}
\newcommand{\AFFkeio}{\affiliation{Department of Physics, Keio University, Yokohama, Kanagawa, 223-8522, Japan}}
\newcommand{\AFFwinnipeg}{\affiliation{Department of Physics, University of Winnipeg, MB R3J 3L8, Canada }}
\newcommand{\AFFkcl}{\affiliation{Department of Physics, King's College London, London, WC2R 2LS, UK }}
\newcommand{\AFFwarwick}{\affiliation{Department of Physics, University of Warwick, Coventry, CV4 7AL, UK }}
\newcommand{\AFFral}{\affiliation{Rutherford Appleton Laboratory, Harwell, Oxford, OX11 0QX, UK }}
\newcommand{\AFFwu}{\affiliation{Faculty of Physics, University of Warsaw, Warsaw, 02-093, Poland }}
\newcommand{\AFFbcit}{\affiliation{Department of Physics, British Columbia Institute of Technology, Burnaby, BC, V5G 3H2, Canada }}
\newcommand{\AFFtohoku}{\affiliation{Department of Physics, Faculty of Science, Tohoku University, Sendai, Miyagi, 980-8578, Japan }}

\AFFicrr
\AFFkashiwa
\AFFicrronly
\AFFmad
\AFFbu
\AFFbcit
\AFFuci
\AFFcsu
\AFFcnm
\AFFduke
\AFFllr
\AFFfukuoka
\AFFgifu
\AFFgist
\AFFuh
\AFFicl
\AFFbari
\AFFnapoli
\AFFpadova
\AFFroma
\AFFkeio
\AFFkek
\AFFkcl
\AFFkobe
\AFFkyoto
\AFFliv
\AFFmiyagi
\AFFnagoya
\AFFkmi
\AFFpol
\AFFsuny
\AFFokayama
\AFFox
\AFFral
\AFFseoul
\AFFsheff
\AFFshizuokasc
\AFFstfc
\AFFskk
\AFFtokai
\AFFtokyo
\AFFtodai
\AFFipmu
\AFFtit
\AFFtus
\AFFtoronto
\AFFtriumf
\AFFtsinghua
\AFFwu
\AFFwarwick
\AFFwinnipeg
\AFFynu

\author{S.~Locke} 
\AFFuci
\author{A.~Coffani}
\AFFllr
\author{K.~Abe}
\AFFicrr
\AFFipmu
\author{C.~Bronner}
\AFFicrr
\author{Y.~Hayato}
\AFFicrr
\AFFipmu
\author{M.~Ikeda}
\author{S.~Imaizumi}
\AFFicrr
\author{H.~Ito}
\AFFicrr 
\author{J.~Kameda}
\AFFicrr
\AFFipmu
\author{Y.~Kataoka}
\AFFicrr
% ...

\date{\today}

\begin{abstract}
Radioactivity induced by cosmic muon spallation is a dominant source of backgrounds for $\mathcal{O}(10)~$MeV neutrino interactions in water Cherenkov detectors. In particular, it is crucial to reduce backgrounds to measure the solar neutrino spectrum and find neutrino interactions from distant supernovae. In this paper we introduce new techniques to locate muon-induced hadronic showers and efficiently reject spallation backgrounds. 
Applying these techniques to the solar neutrino analysis with an exposure of $2790\times22.5$~kton.day increases the signal efficiency by $12.6\%$, approximately corresponding to an additional year of detector running.
Furthermore, we present the first spallation simulation at SK, where we model hadronic interactions using FLUKA. The agreement between the isotope yields and shower pattern in this simulation and in the data gives confidence in the accuracy of this simulation, and thus opens the door to use it to optimize muon spallation removal in new data with gadolinium-enhanced neutron capture detection. 
\end{abstract}

\maketitle

\title{Spallation Paper}
\author{smlocke }
\date{December 2020}

\section{introduction}
Spallation from cosmic-ray muons produces radioactive isotopes and induces one of the largest backgrounds for the Super-Kamiokande (SK) neutrino signal between ${\sim}6$ and ${\sim}25$~MeV. Reducing this background is pivotal for the success of many different analyses in this energy range, and has major implications in solar, reactor, and supernova relic neutrino searches. Specifically in SK, cosmic ray muons and the showers they produce sometimes interact with $^{16}\mbox{O}$ nuclei within the detector volume, producing radioactive isotopes. 

Showers induced by the muons are primarily electromagnetic in nature ($\gamma$-rays and electrons) as a result of delta-ray production, pair production, and bremsstrahlung. However, there is also the possibility for muons to produce secondary particles in the form of neutrons, pions, and others. Recent simulation studies have shown that most spallation isotopes are produced by these secondary particles, with only $11\%$ of isotopes being made directly from muons, and hence these isotopes can be found up to several meters away from the muon track~\cite{BLi_1,BLi_2,BLi_3}. These isotopes then undergo mostly $\beta$ or $\beta\gamma$ decays, mimicking the expected signal for neutrino interactions. Their half-lives extend from milliseconds to seconds, and thus can be much larger than the time interval between two muons in SK, where the muon rate is about $2$~Hz. Identifying spallation isotopes by pairing them with their parent muons is therefore particularly challenging.

In previous SK analyses, spallation reduction algorithms characterized muon signatures solely by considering their reconstructed tracks and prompt light deposition patterns. This method had important limitations due to its reliance on the muon track reconstruction quality. Moreover, while the muon prompt light contains information about produced showers, the shower signatures are partially obscured by muon Cherenkov light. As a consequence, cylindrical cuts around the entire muon track are often necessary.

For the solar neutrino analysis in SK, applying a likelihood cut based on time difference, distance to the muon track and muon light yield  removed $20\%$ of the signal while rejecting $90$\% of the background in the $6.0$-$19.5$~MeV kinetic energy range. The remaining background is dominated by decays of $^{16}\mbox{N}$, which is not only the most abundantly produced isotope~\cite{SKspall_zhang}, but also is particularly difficult to identify \cite{skivsolar}. 

$^{16}\mbox{N}$ is primarily produced through $(n, p)$ interactions on $^{16}$O involving neutrons from muon-induced hadronic showers. Since such neutrons can reach GeV-scale energies,  $^{16}\mbox{N}$ can be found up to several meters away from the muon track. This large distance, together with the long half-life of this isotope, $7.3$~s, makes it particularly difficult to correlate $^{16}\mbox{N}$ decays with their parent muon. Moreover, these decays occur either through the $\beta\gamma$ (66\%) or the $\beta$ (28\%) channel, producing particles with energies ranging between $3.8$ and $10.4$~MeV, well within the solar neutrino energy range. Hence, removing $^{16}$N using only muon track information is particularly difficult and results in significant reduction in neutrino signal efficiency.

Showers producing $^{16}\mbox{N}$ typically contain many neutrons ($\mathcal{O}(100)$) \cite{BLi_comm} which capture on H after thermalizing, as follows:
\begin{equation}
    n + \mathrm{^1}\mbox{H} \rightarrow \mathrm{^{2}}\mbox{H} + \gamma\ (2.2\ \mbox{MeV})
\end{equation}
A single 2.2 MeV $\gamma$ is difficult to be seen in SK, but the large neutron multiplicity makes it possible to directly tag the showers and use them for spallation identification purposes. 

In this paper, we develop a framework to characterize muon-induced spallation processes. In particular we describe new methods to improve spallation identification at SK by tagging the hadronic shower components, identifying clusters of spallation isotopes, and expanding the previously developed spallation cut.  Additionally we present a complete simulation of cosmic muon spallation in SK, inspired by the FLUKA simulations~\cite{fluka_paper,Battistoni:2015epi} developed by~\cite{BLi_1}. 

 The characteristics of the SK detector and its trigger system is described in Sec.~\ref{sec:SKdetector}, and the reconstruction algorithms targeting muons and low energy events are explained in Sec.~\ref{sec:eventreco}. Then, in Sec.~\ref{sec:simulation} we present a FLUKA-based simulation framework modeling muon propagation and shower generation in water.  In Sec.~\ref{sec:neutspall} we then show how to take advantage of a recently deployed trigger system at SK to characterize muon-induced hadronic showers and use them to reduce spallation backgrounds. We demonstrate the ability of our simulation to accurately model the shapes and sizes of these hadronic showers, and thus be used to design future analysis strategies, in Secc~\ref{sec:simucompare}. Finally, in Sec.~\ref{sec:solaranalysis} and Sec.~\ref{sec:spacuts}, we discuss how to use the insight gained through our approach to design new spallation reduction tools for the solar neutrino analysis. We first describe the solar neutrino analysis strategy and the spallation cut used for previous searches. Then we propose an update of this spallation cut and discuss how this new approach leads to a significant increase in the signal acceptance. Although we focus on the solar neutrino analysis for this paper, the techniques we present here can be readily applied to a wide range of low energy neutrino searches, targeting e.~g.~astrophysical transients and the diffuse supernova neutrino background. 
 We measure the yields of spallation isotopes in Sec.~\ref{sec:yields} and demonstrate the FLUKA-based simulation's ability to successfully predict these yields within hadronic model uncertainties.
 Sec.~\ref{sec:conclusion} discusses the paper's conclusions.
\section{Super-Kamiokande Experiment}
\label{sec:SKdetector}

Located in Gifu Prefecture, Japan, the SK experiment is a 50~kton cylindrical water Cherenkov detector, with $2700$~m water equivalent overburden. Major changes in the detector or its electronics define the phases of the experiment~\cite{Super-Kamiokande:2002weg}. This analysis is based on the SK-IV data-taking phase, the longest running phase of SK (August 2008--May 2018). The experiment is composed of an optically separated inner detector (ID) with 11,129 20-inch PMTs and an outer detector (OD) with 1885 8-inch PMTs from Hamamatsu, with the inner detector having a height of 36.2~m and radius of 16.9~m (32.5~kton). The OD acts as a buffer for backgrounds emanating from the surrounding rock and a cosmic-ray veto. The information from the OD is sometimes used to fit and categorize cosmic-rays as well. Inside the ID a fiducial volume (FV) is defined by selecting events with vertices located at least 2~m away from the ID boundary in order to reduce backgrounds arising from radioactivity in the surrounding rock, the PMTs, their stainless steel support structure as well as the tank.

The detector response has been modeled using various calibration procedures, a detailed description of which can be found in~\cite{skcalib}. The two sources most relevant for low energy analyses are a linear accelerator~(LINAC) and a
Deuterium-Tritium~(DT) neutron generator. 
The LINAC calibration checks the absolute energy scale as well as vertex and direction reconstruction, by injecting mono-energetic electrons into SK at various locations, as described in~\cite{16Ncalibrationsource}. 
The DT generator produces 14~MeV neutrons from DT fusion; they in turn produce $^{16}$N isotopes from $\mathrm{^{16}O}$ via $(n, p)$ reactions, as described in~\cite{16Ncalibrationsource}. The $^{16}$N decays are used to check the absolute energy scale. The DT calibration is much easier to perform compared to LINAC calibration and therefore the DT calibrations are done at regular time intervals while LINAC calibrations are only done once every few years. Also, LINAC electrons are limited to the downward direction while the $^{16}$N decays isotropically.
\label{sec:softtrig}

SK-IV introduced QTC Based Electronics with Ethernet (QBEE) and new data acquisition (DAQ) computers allowing for enhanced data processing capability~\cite{Nishino:2009zu}. The new DAQ substantially increase the bandwidth of the information collected from the PMTs: each PMT triggers by itself and then its integrated charge and trigger time is digitized and processed. The QBEEs are equipped with three different amplifiers for the charge digitization. These amplifiers enable a dynamic charge detection range, allowing $\sim$5 times increase in the maximum detected charge for an individual PMT. 

Due to these upgrades the former hardware trigger is replaced by software triggers: Data are acquired in 17~$\mu$s segments (hardware triggers) controlled by a 60~kHz clock. The 17~$\mu$s segments connect seamlessly to each other. The standard software trigger applies a simple coincidence criterion: at least 31 triggered PMTs (or hits) within about 200~ns of each other. The dark noise rate per 200~ns of all PMTs is about 11, so a coincident signal from about 20~PMTs is needed. PMT data are saved in a window starting from 500~ns before and 1000~ns after trigger time. If the coincidence is larger than 47 triggered PMTs, the trigger window is expanded to stretch from 5~$\mu$s before and 35~$\mu$s after the trigger time. While the standard SK trigger is well-suited for events with kinetic energies down to $\sim$3.5~MeV, it is not sensitive to neutron captures on hydrogen which are the key to identifying hadronic showers in SK-IV.

A neutron capture on hydrogen emits a single 2.2~MeV $\gamma$-ray. Such events produce little Cherenkov light; in SK-IV 2.2~MeV $\gamma$s result in only about seven detected photo-electrons on average, so the standard software trigger efficiency is very small. After an event of at least $\sim$8~MeV electron-equivalent energy, the standard software trigger will automatically issue after-triggers (AFT) that record all hit PMTs within the next 500~$\mu$s. A cosmic muon easily fulfills this condition and the AFT triggers would catch most 2.2~MeV $\gamma$-rays from subsequent neutron captures, although reliable identification of the 2.2~MeV $\gamma$ signal is possible for only $\sim$20\% of them. However, to not unduly strain the standard software trigger, these AFT triggers are disabled after cosmic muons and thus do not allow to identify neutrons from muon-induced showers. 

The Wideband Intelligent Trigger (WIT) receives the full copy of all PMT data before triggering and runs in parallel to the standard software trigger~\cite{wit,Elnimr:2017nzi}. It is designed to trigger on electrons of at least 2.5~MeV kinetic energy and has some efficiency to stand-alone triggering of 2.2~MeV $\gamma$-rays. WIT is implemented on a computer farm consisting of ten machines connected via 10~GBit Ethernet lines. Seven machines are dual-CPU with eight cores per CPU (16~hyperthreaded cores per CPU) while three newer machines have 28~cores per CPU (56~hyperthreaded cores). WIT receives blocks of 1344~consecutive hardware triggers ($\sim$23~ms). These blocks overlap by 64 hardware triggers (1.1~ms) and are distributed among nine online machines where they are processed separately in one of the hyperthreaded cores. The remaining one of the ten computers sorts all processed files and assembles ``subruns": 4,000~consecutive blocks gathered into one file ($\sim$90~s of data). All data processing and sorting is handled without disk write operations, fast RAM memory of the WIT computers replaces the usual hard drives.

The trigger criterion of WIT is more complex than the standard software trigger as coincidence is applied to PMT hit time residuals
$\Delta t_i$ with respect to a list of possible vertices $\vec{v}_\alpha$:
$\Delta t_i=t_i-|\vec{v}_\alpha-\vec{p}_i|/c_{\mbox{\tiny water}}-t_0$ with the PMT hit vectors $\vec{p}_i$ and the light emission time $t_0$ and the speed of light in water $c_{water}$.
The applied trigger condition is:
\[
sg=\mbox{Max} \sum_{\mbox{\tiny all PMT hits } i} e^{-\frac{1}{2}\left( \frac{\Delta t_i}{\sigma}\right)^2}>6.6.
\]
The time uncertainty is chosen to be $\sigma=5$~ns. ``Max'' refers to the list of possible vertices. In order to create this list, four hit combinations among a set of selected hit PMTs are chosen, and these four hit time residuals are required to be exactly zero in order to define a possible vertex. To improve the speed of the algorithm, a pre-trigger condition of $>$~dark noise~$+11$ hit coincidence within an absolute time window of 230~ns is applied, and the hits in that window are required to obey relations $\delta x_{ij}>c_{\mbox{\tiny water}} \delta t_{ij}$ here $\delta t_{ij}$ ($\delta x_{ij}$) are the time (spatial) difference between hit $i$ and $j$. The raw PMT times and pulse heights are converted from the digitized counts to calibrated times and photo-electrons in real time.

If a trigger is found, a fast vertex fit to the set of hits used for the construction of four-hit combinations is done. Only events reconstructing further than 1.5~m from any PMT are passed to the online version of the standard SK vertex fitter. For that, a 1.5~$\mu$s window is formed around the trigger time ($500$~ns before and $1000$~ns after). If the event reconstructs at least 2~m from any PMT, and if the number of hits with time residuals between $-6$~ns and $+12~$ns is larger than ten, the event is saved. The trigger efficiency of 2.2~MeV $\gamma$-rays is 13\% averaged over the entire detector and 17\% averaged over the fiducial volume.

The $1.5~\mu$s events are stored in ROOT format output files \cite{root} containing raw PMT time and pulse height digitized counts, calibrated PMT times and pulse heights, the trigger time and position, the reconstructed vertices by both vertex fits, and the the number of PMT hits within 18~ns. Also stored are GPS time stamps and pedestal and calibration measurements. Storing these low-energy events will allow the resolution of the structure of muon-induced hadronic showers and the design of an efficient spallation reduction strategy at SK-IV.
\section{Event Reconstruction}
\label{sec:eventreco}

\subsection{Cosmic Rays and Track Fits}
\label{sec:muons}
Muons pass through SK at a rate of approximately 2~Hz, and their tracks are reconstructed from the PMTs hit within the detector. The muon reconstruction used for this analysis (outlined in \cite{muboy1, muboy2}) accurately fits tracks as well as categorizing muons. After removing PMT noise hits, the fitter makes an initial guess on the track using the earliest hit PMT with at least three neighboring hits as an entry point and time, and the largest cluster of charge as the exit point. The track parameters are then varied and a likelihood dependent on the expected Cherenkov light pattern is maximized to get a final track fit. Muons are categorized based on characteristics of the observed light, with the four different muon categories described by:
\begin{enumerate}
    \item Single Through Going ($\sim$82\%): Single muons that pass entirely through the detector and is the default fit category for a muon. 
    \item Stopping ($\sim$7\%): Single muons that enter the ID but do not exit it. Identified by low light observed near the exit point of the muon and nearby OD information. 
    \item Multiple ($\sim$7\%): Bundles of muons passing through the detector simultaneously. Identified by light inconsistent with a single Cherenkov cone. 
    \item Corner Clipping ($\sim$4\%): Single through going muons found to have a track length of less than 7~m inside the ID, while also occurring near the top or bottom of the detector. 
\end{enumerate}
To check the fitter accuracy, $\sim$2000~events were fit by this method and by hand. For the categories found by the fitter, $\sim$0.5\% of single through going, $\sim$1.4\% of corner clipping, $\sim$13\% of multiple, and $\sim$30\% of stopping muons were found to be something else by eye-scan. The difference in the mistagged stopping muons were hand fit to be through going muons. Resolution studies found the entry point resolution to be 100~cm for all types of muons, except multiple muons with more than 3 tracks, and a directional resolution of 6$^\circ$. In this analysis, if multiple tracks were fit, the principal track is used to identify subsequent events. 
\subsection{Reconstruction of low energy events}
\label{sec:lowereco}
Events are reconstructed from their hit timing (vertex), hit pattern (direction), and the effective number of hits observed (energy). Events below 19.5 MeV only travel several cm within SK and are treated as a point source. The vertex is reconstructed by maximizing a likelihood dependent on the timing residuals of the hit PMTs, $\tau_i$:
\begin{equation}
    \tau_i = t_i - t_{TOF} - t_0
\label{eq:timeres}
\end{equation}

where $t_i$ is the time of the PMT hit, $t_{TOF}$ is the time of flight from the test vertex to the PMT, and $t_0$ is the event time. This likelihood is unbinned, and therefore it cannot be used to evaluate the goodness of the vertex fit~($g_t$). 

To create a goodness of fit~($g_t$) from the timing residuals $\tau_i$ the weighted sum of Gaussian functions $G(\tau_i|\sigma)=\exp\left[-0.5\left(\tau_i/\sigma\right)^2\right]$ is used:
\begin{equation}
    g_t=\sum_i^{N_{hit}}W_i G(\tau_i|\sigma)
\label{eq:gvt}
\end{equation}
where $\sigma=5$~ns is the width of the Gaussian, obtained by combining the single photoelectron PMT timing resolution of 3~ns to the effective time smearing due to light scattering and reflection. The weights $W_i$ are $W_i=G(\tau_i|\omega)/(\sum_j G(\tau_j|\omega))$
with a ``weight width" of $\omega=60$~ns.

Event direction reconstruction is a maximum likelihood method comparing data and MC simulation of the PMT hit pattern caused by Cherenkov cones. This likelihood is dependent on reconstructed energy and the angle between the event direction and the direction to individual PMTs. Using the reconstructed direction, the azimuthal symmetry of the PMT hit pattern is probed with the goodness $g_p$, a KS test:
\begin{equation}
\begin{gathered}
g_p = \frac{\max[\phi^{uni}_i-\phi^{data}_i] - \min[\phi^{uni}_i-\phi^{data}_i]}{2\pi}, \\
\phi^{uni}_i = \frac{2\pi\cdot i}{N_{hit}},
\end{gathered}
\label{eq:gdir}
\end{equation}

\noindent where $\phi^{uni}$ is the angle for evenly spaced hits around the event direction and $\phi^{data}$ is the actual hit angle around event direction. Like $g_t$, $g_p$ also tests the quality of the vertex reconstruction: a badly misplaced vertex often presents the direction fitter with a Cherenkov cone pattern appearing too small (or too big), which implies an accumulation of hit PMTs on ``one side" of the best-fit direction.

Finally, for the energy reconstruction, we evaluate the photons' times of flight from the reconstructed vertex to the hit PMTs and subtract them from the measured arrival times. We then define the effective number of hits $N_{\mbox{\tiny eff}}$ as the maximal number of hits in a 50~ns coincidence window.
This number is then corrected for water transparency, the angle of incidence to the PMTs and photocathode coverage, dark noise rate, PMT gain over the course of SK-IV, PMT occupancy around a hit, the PMT quantum efficiency, and the fraction of live PMTs. The energy is then calculated from a fifth order polynomial dependent on $N_{\mbox{\tiny eff}}$ for energies in the solar neutrino range. The energy reconstruction assumes an electron interaction. This is important to note as neutron captures on hydrogen have a single $\gamma$ which creates less light than a 2.2 MeV electron would. Within this paper the energy of events will be given in terms of the kinetic energy of an electron with equivalent light yield.

\section{Spallation simulation}
\label{sec:simulation}

\subsection{Cosmic muon simulation}
\begin{figure}
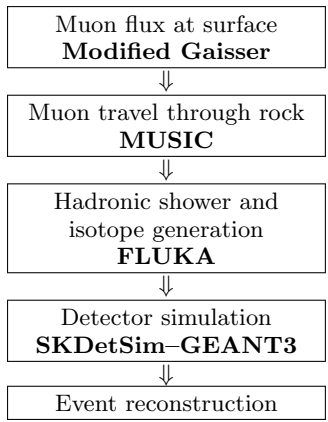

    \centering
    \fbox{\begin{minipage}{4cm}
    Muon flux at surface\\
    {\bf Modified Gaisser}
    \end{minipage}}\\
    $\Downarrow$\\
    \fbox{\begin{minipage}{4cm}
    Muon travel through rock\\
    {\bf MUSIC}
    \end{minipage}}\\
    $\Downarrow$\\
    \fbox{\begin{minipage}{4cm}
    Hadronic shower and isotope generation\\
    {\bf FLUKA}
    \end{minipage}}\\
    $\Downarrow$\\
    \fbox{\begin{minipage}{4cm}
    Detector simulation\\
    {\bf SKDetSim--GEANT3}
    \end{minipage}}\\
    $\Downarrow$\\
    \fbox{\begin{minipage}{4cm}
    Event reconstruction
    \end{minipage}}\\
    \caption{Simulation steps, from the modeling of the muon flux at the surface to event reconstruction in SK.}
    \label{fig:simupipeline}
\end{figure}

In order to understand and optimize spallation event removal techniques we simulate the interactions of cosmic muons and the subsequent production of neutrons and isotopes in the SK water. 
This muon simulation is composed of five parts. We first model the muon flux at the surface of the Earth using a modified Gaisser parameterization described in Ref. \cite{muonsimulation}, and propagate muons through the rock to SK using a dedicated transport simulation code.
Second, we simulate the production of hadronic showers and radioactive isotopes inside SK using FLUKA \cite{fluka_manual, fluka_paper}. The FLUKA results are then injected into SKDetSim, the official GEANT-3 \cite{geant3} based detector simulation for SK, that will model detector effects as well as minimum ionization around the muon track. Finally, we reconstruct muon tracks, neutron captures, and isotope decays using standard SK reconstruction software as well as the procedure described in Sec.~\ref{sec:neutspall}. The simulation pipeline is summarized in Fig.~\ref{fig:simupipeline}.
\subsubsection{Muon generation and travel}
Most of the muons reaching the Earth's surface are produced at an altitude of around $15$~km, from the interactions of primary cosmic rays in the atmosphere, principally the decays of charged mesons~\cite{gaisser_book}. The shape of the meson production energy and angular distributions reflects a convolution of the production spectra, the energy loss, and the decay probability in the atmosphere. In this study we model the muon flux at the surface using a modified Gaisser parameterization, that has been optimized for detectors at shallow depth such as SK~\cite{muonsimulation}. While muons contributing most to spallation background can cross the detector either alone or as part of muon ``bundles'' --caused by meson decays within cosmic ray showers in the atmosphere-- differences in spallation observables between these two configurations will be entirely due to track reconstruction issues. In this paper, our primary goal is to evaluate FLUKA's ability to model muon-induced showers and isotope production in SK. We will therefore consider only single muons and will leave the subject of bundles to a future study.

In order to obtain the muon flux entering the SK detector we now need to propagate muons through the rock surrounding the detector. We simulate muon propagation using the MUSIC \cite{music,music2,music3} propagation code. MUSIC integrates models for all the different types of muon interactions with matter leading to energy losses and deflections, such as pair production, bremsstrahlung, ionization and muon-nucleus inelastic scattering. Angular and lateral displacements due to multiple scattering are also taken into account. Muons are transported with energies up to 10$^7$ GeV. Here, we used the rock composition model described in~\cite{muonsimulation}, with an average density of $\rho = 2.70~\mathrm{g\,cm^{-3}}$. We compute the muon travel distance within the rock as a function of the incident angle using a topological map of the SK area from 1997~\cite{muonsimuKAMLAND,geoinstitute}.

\begin{figure}[tbph]
	\centering
	\includegraphics[width=\linewidth]{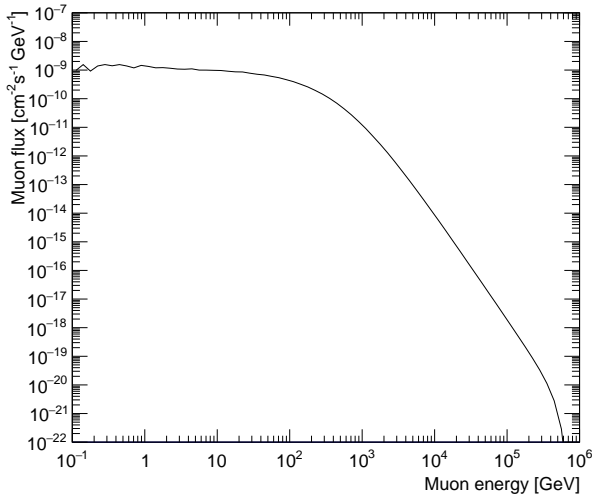}
	\caption{Muon energy spectrum at the location of SK detector in the mine inside Mt. Ikenoyama.}
	\label{fig:musi_final_distrib}
\end{figure}

Figure~\ref{fig:musi_final_distrib} shows the energy spectrum of the muons that reach SK. The rock above the detector constitutes a particularly efficient shield, effectively blocking muons with energies lower than $600$~GeV. The cosmic muon flux is reduced from $6.5\times10^{5}~\mu \mathrm{m^{-2}\,h^{-1}}$~\cite{borexino} to $1.54 \times 10^{-7}~\mathrm{cm^{-2}\,s^{-1}}$ which corresponds to a muon rate of 1.87~Hz, as expected from previous measurements~\cite{muonrate1, muonrate2, muonrate3}.

\begin{figure}
    \centering
    \includegraphics[width=\linewidth]{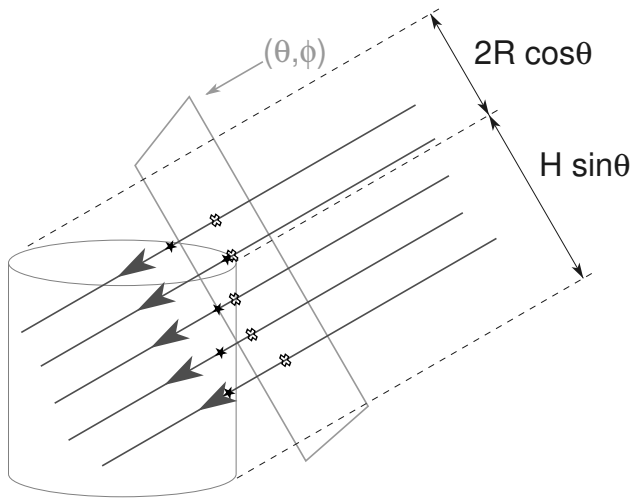}  
    \caption{Spatial distribution of trajectories for muons produced in the same area of the sky. These muons can be considered almost parallel when reaching SK and the intersection of their trajectories with a plane perpendicular to their direction will be uniformly distributed. Here, $R$ and $H$ are the radius and the total height of SK's inner detector while $\theta$ and $\phi$ define the direction of the muons. Here, the stars indicate the muon entry points and the crosses indicate the intersections of the muon trajectories with the plane.}
    \label{fig:muentrypoints}
\end{figure}

Although MUSIC allows to evaluate the effect of muon transport through rock on the muon directional distribution, it does not account for the detector's cylindrical geometry. We account for these effects by assuming that cosmic muons are isotropically produced in the atmosphere and that all muons produced in the same area of the sky have quasi-parallel trajectories inside SK. For each muon generated by MUSIC with a given direction $(\theta,\phi)$, with origin of the coordinates at the center of SK, $\theta$ representing the zenith angle and $\phi$ the azimuthal one set to zero when the final muon travels from east to west, we generate a set of parallel tracks with uniformly distributed intersection points in the plane perpendicular to the $(\theta,\phi)$ vector, as shown in Fig.~\ref{fig:muentrypoints}. We then reject all the tracks that do not cross the detector, thus straightforwardly accounting for geometrical effects. This procedure allows us to convert the directions generated by MUSIC into a sample of entry points distributed on the surface of the SK inner detector. 
\subsubsection{Muon interactions in water}
Propagation and interactions of muons in water are simulated with FLUKA, taking as input MUSIC energy and angular distributions. FLUKA \cite{fluka_manual, fluka_paper} is a general purpose Monte Carlo code for the description of interactions and transport of particles in matter. It simulates hadrons, ions, and electromagnetic particles, from few keV to cosmic ray energies. It is built and frequently upgraded with the aim of maintaining implementations and improvements of modern physical models. FLUKA version 2011.2x.7 is used for this work, together with FLAIR (version 2.3-0), an advanced user interface to facilitate the editing of FLUKA input files, execution of the code and visualization of the output files \cite{flair}. FLUKA propagates muons into the SK detector, simulating all the relevant physics processes that lead to energy losses and creation of secondary particles: ionization and bremsstrahlung, gamma-ray pair production, Compton scattering and muon photonuclear interactions.  Hadronic processes such as pion production and interactions, low energy neutron interactions with nuclei and photo-desintegration are also modeled.

FLUKA code fully integrates the most relevant physics models and libraries. For this work, the simulation was built with the default setting PRECISIO(n). All the specifics related to this setting can be found in \cite{fluka_manual}.  More detail about the models and settings used in this paper can be found in appendix~\ref{appendix_sim}. In particular, low-energy neutrons, which are defined to have less than 20~MeV energy, are transported down to thermal energies, a setting that is critical for our study. 

Crucial options complement the default setting: EVAPORAT(ion) and COALESCE(nce) give a detailed treatment of nuclear de-excitations while nucleus-nucleus interactions are enabled for all energies via the option IONTRANS.
\vspace{.3cm}\\
The SK detector is modeled as a cylindric volume of pure water, as described in Sec.~\ref{sec:SKdetector}. The PMT structure is not simulated in FLUKA, given that it is fully incorporated in SKDetSim. Since muons can induce showers outside the water tank and secondary products may reach the active part of the detector, previous studies~\cite{BLi_1} examined the effect of a 2~m thickness of rock surrounding the OD:  it was proven that this has a minor effect on the results. Thus the rock,  as well as the tank and the support structure, are not simulated in this work. Both negative and positive muons are generated assuming a muon charge ratio, defined as the number of positive over negative charged muons, of $N_{\mu^+}/N_{\mu^-}$ = 1.27 \cite{CMSchargeratio}. Note that the measured values of the charge ratio at SK depth  can vary by about 20\%, with the highest value (1.37$\pm$0.06) measured at Kamiokande~\cite{Yamada:1991aq}. However, since the isotope yield depends only weakly on the muon charge, these variations have a negligible impact on this analysis, with an effect on the predicted yields of less than 1\%.

\subsubsection{Detector response and event reconstruction}

We model the detector response using the official detector simulation for Super-Kamiokande, referred to as SKDetSim. This simulation is based on GEANT 3.21~\cite{geant3} for detector modeling and uses a customized model for light propagation. It covers all aspects of event detection, from the initial interaction to the light collection on the PMTs and event reconstruction.

For this purpose, SKDetSim models in detail the entire geometry of the detector, the particle propagation in water, the emission of Cherenkov photons, reflection and absorption of light on materials, photo-electron production, and electronic response. Simulated data and real data are processed similarly. For this reason, particles simulated in SKDetSim are recorded in trigger windows corresponding to the ones applied to data. For each event, the detector dark rate is also simulated and can be added or not to the outputs depending on the specific needs.

Since FLUKA already models the shower development in water, special care must be taken when interfacing with SKDetSim, which is used for the detector response, including light propagation and collection. It is important to avoid SKDetSim doing a parallel generation of muon-induced showers, which would lead to a double counting of isotopes. In this study, we therefore focus on observables that are particularly robust against possible mismodeling of the shower Cherenkov light pattern: the yields of the spallation-produced isotopes, and the characteristic neutron clouds discussed in Sec.~\ref{sec:hadronic_spall}. Isotope decays and neutron captures can indeed be simulated in isolation from their parent muons in SKDetSim, and will hence not lead to unwanted interactions. Muon-induced showers still need to be simulated, as they affect the reconstruction of the muon track and hence neutron identification; however this effect on reconstruction is limited. The impact of shower mismodeling can therefore be mitigated using a few simple steps. The following points describe the interfacing between FLUKA and SKDetSim for a typical event, which consists in a  primary muon, a hadronic shower, potential neutron captures and isotopes decays.

An essential change in the input card is the deactivation of muon-nucleus interactions (GEANT-MUNU is set to zero) to prevent the muon from inducing extra showers. 
Thus, the muon will averagely behave only as an ionizing particle. Radiative losses through interactions with atomic nuclei are deactivated but processes such as bremsstrahlung and pair production are still possible and their importance increases with muon energy.

Together with the muon, particles produced in hadronic and electromagnetic showers in FLUKA are injected at $t = d/c_{\mbox{\tiny vac}}$, where $d$ is the distance from the muon entry point to the shower particle. This typically corresponds to a few nanoseconds. 
Here again, in order to avoid double-counting, we deactivate photofission as well as secondary particle generation for all hadronic interactions in SKDetSim, that is used only to model electromagnetic~(EM) processes, including the emission of Cherenkov light. Conversely, we only inject particles from FLUKA that are commonly produced in hadronic showers and can lead to the prompt emission of Cherenkov light, that is, $\gamma$-rays, pions, and kaons. The only exception occurs if a radioactive isotope is produced in a shower initiated by electromagnetic processes, e.g. by interactions involving electrons, photons, or positrons. In this case, the particles initiating the shower are also injected into SKDetSim, provided their energy is larger than 0.1~GeV since no isotope production has been observed below this threshold in previous simulation studies~\cite{BLi_2}. This scenario is however particularly rare and the impact of these extra particles on the muon light pattern will be extremely limited. Table \ref{tab:processes_FLUKA_SKDETSIM} summarizes the processes treated in FLUKA and SKDetSim respectively. 
\begin{table}
	\begin{center}
		\caption{\label{tab:processes_FLUKA_SKDETSIM}The Table summarizes the main processes activated and deactivated in SKDetSim and the particles generated or ignored in FLUKA and injected in SKDetSim. More description in the text.}
		\begin{tabular}{cc}

			\toprule
			\multicolumn{2}{c@{\quad}}{SKDetSim} \\
			Deactivated & Activated    \\
			\hline
			Photofission & EM interactions \\
			Secondary part.  & Cherenkov radiation  \\
			generation in inel. int.& \\
			& Decays \\
	
			\toprule
			\multicolumn{2}{c@{\quad}}{FLUKA}\\
		     Generated & Ignored      \\
			\hline
			 EM show. w/ spall. & EM show. w/o spall. \\
			 (if $E>0.1$ GeV) &  \\
		 $\gamma, \pi, K$ from inel. int. &  \\
			 $\gamma$ from $n$ captures &\\
			 Isotope decay prod. & \\
			\bottomrule        
			
		\end{tabular}
	\end{center}
\end{table}

While the Cherenkov light emission from pions and $\gamma$-rays coincides with the one from muon ionization, neutron capture typically occurs over much larger time scales. For a given muon, we therefore treat each neutron capture separately in SKDetSim. Since neutrons have already been propagated by FLUKA, we directly simulate a $2.2$~MeV $\gamma$-ray at each capture vertex.

Finally, using the isotope production vertices and decay times given by FLUKA, we simulate isotope decay products in SKDetSim.

Each isotope decay can produce either a $\gamma$ or a $\gamma$- and a $\beta$-ray, sometimes followed by a neutron. In what follows we will consider only products leading to prompt Cherenkov light emission, as their reconstructed energy distribution will affect isotope yield measurements.

Each of the steps described above requires modeling not only the signal, but also the noise in the detector. For muons, shower particles, and isotope decays, we use the modeling of the PMT dark noise from SKDetSim, based on regular measurements made over the SK-IV period. The treatment of neutron capture signals is more complex; this signal in water is particularly weak and the predicted performance of the neutron tagging algorithm is highly sensitive to dark noise modeling. Therefore, we inject noise samples from data into the signal simulation results. These samples were taken at different times over the whole SK-IV using a random trigger, and can hence be used to reflect the time variations of the noise in the detector. Finally, the muon, neutron capture, and radioactivity events undergo the same reconstruction and reduction steps as the ones described in Sec.~\ref{sec:SKdetector} and \ref{sec:hadronic_spall} for data. WIT triggers were used to reconstruct neutron captures.

\section{Spallation Tagging with Hadronic Showers}
\label{sec:neutspall}
This section describes how to identify neutron clouds associated with the hadronic showers induced by cosmic muons, and exploit their space and time correlations with spallation isotopes to reduce spallation backgrounds. Efficient reduction requires not only reliably identifying neutrons but also accurately reconstructing their positions in order to locate the shower and characterize its shape. As discussed in Sec.~\ref{sec:softtrig}, however, neutrons produced in SK-IV can be identified only through their capture signal on hydrogen: a single $2.2$~MeV $\gamma$-ray. Since this signal is below the threshold of standard triggers and the AFT trigger is disabled for OD signals~(all muons except for fully contained atmospheric $\nu_{\mu}$ CC interactions), we use the WIT trigger described in Sec.~\ref{sec:softtrig} to select neutron candidates. In this section, we discuss the relevant observables needed to refine this initial selection and reliably characterize neutron clouds, and present several case studies illustrating the validity of our approach. In Sec.~\ref{sec:simucompare}, the characteristics of the observed neutron cloud will be compared to predictions from the simulation described in Sec.~\ref{sec:simulation}. Neutron identification will additionally be used to build spallation-rich samples and refine the computation of the spallation isotope yields at SK in Sec.~\ref{sec:yields}. Finally, an example of incorporating neutron cloud cuts into a full spallation analysis at SK will be shown in Sec.~\ref{sec:spacuts}.

\subsection{Observables for neutron cloud identification}
\label{sec:vardef}
Locating and characterizing neutron clouds requires applying quality cuts on the neutron candidates selected using WIT. These cuts are necessary to select well-reconstructed neutron capture vertices, as well as to remove events associated with, e.~g.~radioactivity, flashing PMTs, and dark noise fluctuations. To this end, we consider two categories of observables, based either on the neutron candidate hit time and light pattern, or on the space and time correlations between neutron candidates and muons. 

\subsubsection{Neutron candidate hit time and light pattern}
The amount of Cherenkov light deposited by an event in SK, as well as the shape of the associated ring, allows to identify well-reconstructed neutron capture events. Using the tools outlined in Sec.~\ref{sec:eventreco}, cuts are placed on neutron capture candidates based on these variables to increase sample purity. 

\paragraph{Reconstructed vertex:} intrinsic radioactivity (except for radon daughters) events typically occur on the walls of the detector or in the surrounding rock. Hence, their reconstructed vertices will typically lie either outside the ID or near its walls. In what follows we will therefore require the reconstructed vertices of the candidate neutrons to lie inside the ID. Note that for events with energies lower than ${\sim}3.5$~MeV the WIT trigger already requires online reconstructed vertices to lie in the FV. WIT triggers above that energy only need to reconstruct inside the detector.

\paragraph{Reconstructed energy $E_{rec}$:} when a neutron is captured on hydrogen, a single 2.2~MeV $\gamma$ is released. The SK energy reconstruction for this analysis assumes an interaction resulting in an electron. Thus, the reconstructed energy is the equivalent to that of an electron interaction with the same amount of Cherenkov light production. Since the energy is calculated based on the number of effective hit PMTs, we will typically require $E_{rec} < 5-6$~MeV to account for fluctuations in the light yield. 

\begin{figure*}
    \centering
    \includegraphics[width=\textwidth]{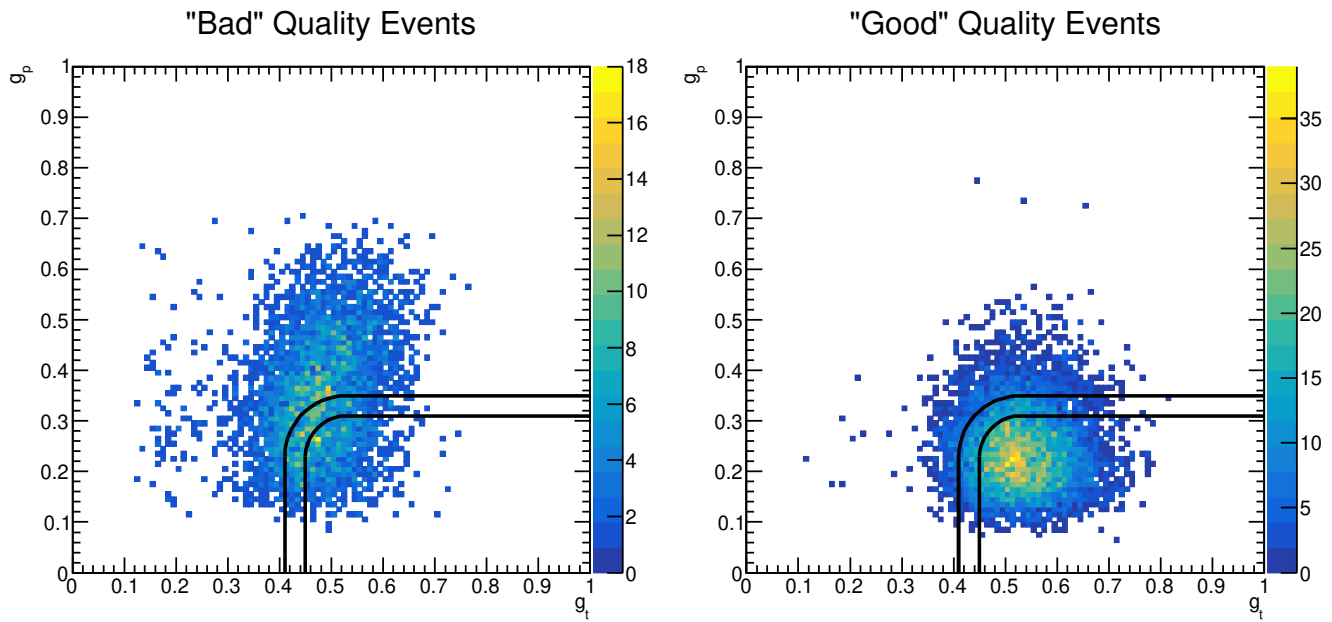}
    \caption{MC generated 2.2~MeV $\gamma$-rays processed by WIT software. The left distribution shows ``poorly" reconstructed events as defined as reconstructed vertex being more then 5~m from MC truth and the right distribution shows the ``good" reconstructions. The lines separate the different weight regions used to parameterize the neutron cloud cuts described in Sec.~\ref{sec:cloudcut}.}
    \label{fig:mcovaq}
\end{figure*}

\paragraph{Reconstruction goodness $g_t$, $g_p$: } these observables have been defined in Sec.~\ref{sec:lowereco} and measure the goodness of the position and direction reconstruction of the neutron candidate. In order to determine suitable spallation cuts, 2.2~MeV $\gamma$-rays were generated using SKDetSim. The events were simulated without dark noise, overlaid with real SK online raw data, and then processed with WIT. The resulting sample was then split into ``good'' and ``bad'' events, whose reconstructed vertices lie less and more than $5$~m away from the true vertices respectively. Figure~\ref{fig:mcovaq} shows the two-dimensional distributions of $g_t$ and $g_p$ for both event categories, with ``good'' events being associated with higher goodness values. In the simple case studies presented in this section, we will only require $g_t > 0.5$. In the more refined strategy discussed in Sec.~\ref{sec:spacuts} we will impose cuts on both $g_t$ and $g_p$ in order to improve the neutron cloud localization. The resulting refined cut is overlaid on the goodness distributions shown in Fig.~\ref{fig:mcovaq}.
 
\subsubsection{Correlations with muons}
\label{sec:corrmu}
Neutrons produced in muon-induced showers will be found close in time and space to muon tracks. The observables we describe here allow to identify these space and time correlations. Moving forward we make the reasonable assumption that muons follow a straight path inside the detector. We then parameterize their tracks using an entry point and a direction, that are determined using the fitter described in Sec.~\ref{sec:muons}. We then define the following observables: the time difference $\Delta t$ between a neutron candidate and a muon, and the transverse and longitudinal (with respect to the muon entry point) distances from the candidate to the muon track, $l_t$ and $l_{\mbox{\tiny \it LONG}}$.

\begin{figure}
    \centering
    \includegraphics[width=\linewidth]{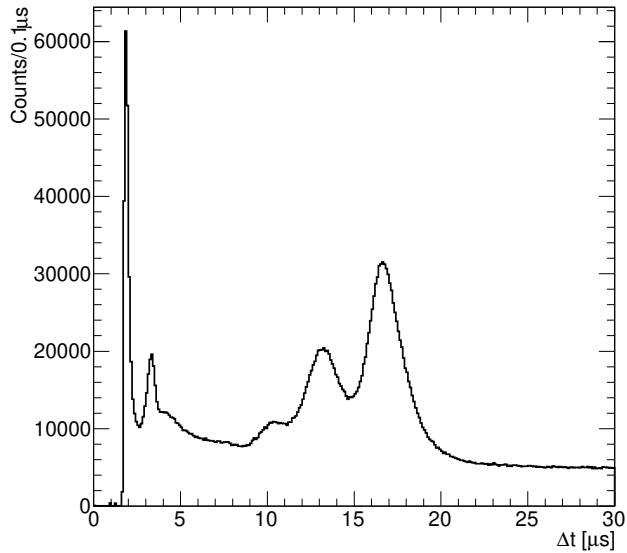}
    \caption{Time distribution of WIT-triggered events with $l_t < 5$~m. The goodness cuts included in the WIT trigger lead to a deficit of events in the 5--10~$\mu$s time range while the bumps are contributions from fake neutrons associated with PMT afterpulsing. In order to remove these fake neutrons, quality events are required to have $\Delta t > 20~\mu$s.}
    \label{fig:afterpulsing}
\end{figure}

\paragraph{Time difference $\Delta t$:} the characteristic neutron capture time in pure water is $\tau_{cap}{\sim}205~\mu$s. We hence expect to find most neutrons within about $500~\mu$s from their parent muon crossing time. Since the muon rate in SK is about $2$~Hz, this time window alone allows to unambiguously link a neutron capture event to its parent muon. Additionally, for high-quality neutron cloud samples, we will require $\Delta t$ to be larger than 20~$\mu$s to account for PMT afterpulsing \footnote{Muons creating hadronic showers also deposit a significant amount of light within the detector, and ionization of residual air within the PMT creates a delayed pulse after drifting to the principal dynode. Although there is a relatively small chance for a PMT to experience this, bright muons have hits at most, if not all PMTs, within the detector and these afterpulsings cause problems. The extra hits create false/poorly reconstructed events in the 10--20~$\mu$s time range and this time region is therefore avoided.} The afterpulsing features are shown in Fig.~\ref{fig:afterpulsing} for events found within 5~m of a muon track.

\begin{figure}
    \centering
    \includegraphics[width=0.57\linewidth]{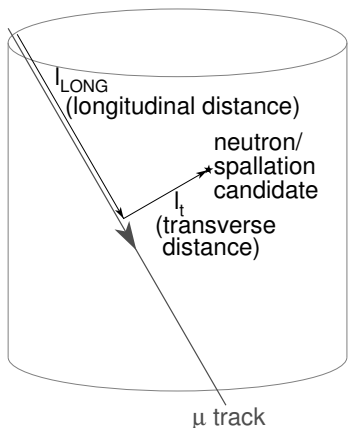} 
    \caption{Diagram showing the $l_t$ and $l_{\mbox{\tiny LONG}}$ observables associated with neutron identification, as described in Sec.~\ref{sec:corrmu}.}
    \label{fig:vardia}
\end{figure}

\begin{figure}
    \centering
    \includegraphics[width=\linewidth]{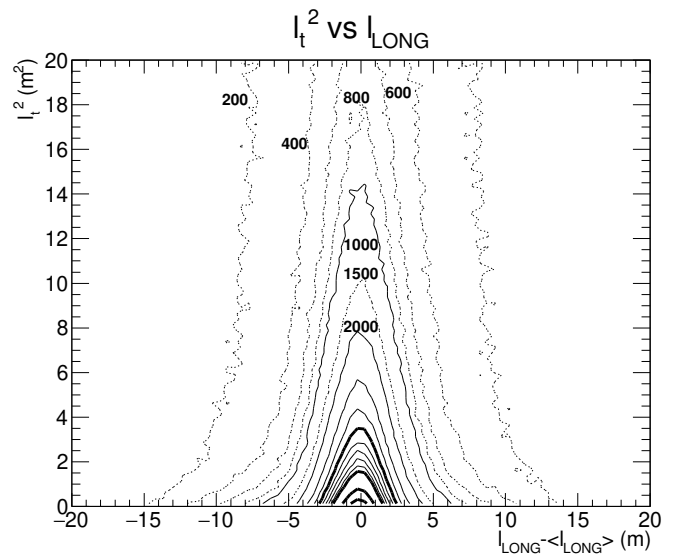}    
    \caption{Data Vertex correlation of neutron capture events. The vertical axis shows the squared distance to the muon track, the vertical axis the distance along the muon track with respect to the average. Solid contours indicate levels of multiple 1,000 events/bin, the thick contours are for 5,000, 10,000, 15,000, and 20,000. The levels of the dotted line contours are marked in the Figure.}
    \label{fig:ltvln}
\end{figure}

\paragraph{Transverse and longitudinal distances, $l_t$ and $l_{\mbox{\tiny \it LONG}}$: }  $l_t$ defines the distance between a neutron candidate and the closest point on the muon track while $l_{\mbox{\tiny LONG}}$ is the longitudinal distance, the distance between that closest point on the muon track and the muon track entry point. The definitions of these two observables are shown in Fig.~\ref{fig:vardia}. Their two-dimensional distribution, shown in Fig.~\ref{fig:ltvln}, allows to characterize the shapes and sizes of the neutron clouds. In this figure, $\langle l_{\mbox{\tiny LONG}}\rangle$ refers to the average $l_{\mbox{\tiny LONG}}$ of all the neutrons in the cloud. Note that neutron clouds have an elongated shape along the muon track and can extend up to about 5~m. Here, we showed $l_t^2$ instead of $l_t$ to reflect the amount of phase space available.


Incorporating the observables defined above in spallation analyses thus allow to identify neutrons and accurately locate neutron clouds and the showers that generated them. In what follows we will show case studies highlighting the validity of our neutron identification and localization procedure. 

\subsection{Neutron cloud identification: case studies}
\label{sec:hadronic_data}
Here, we present examples of how to identify neutrons and reconstruct clouds using the observables defined above. We first demonstrate our ability to identify individual neutrons by associating the WIT trigger with simple goodness and position cuts. Then we use well-reconstructed neutron candidates to build neutron clouds, and exploit their spatial correlation with spallation isotopes.

\subsubsection{Identifying individual neutrons}
\label{subsec:hadronic_data}
\begin{figure}
    \centering
    \includegraphics[width=\linewidth]{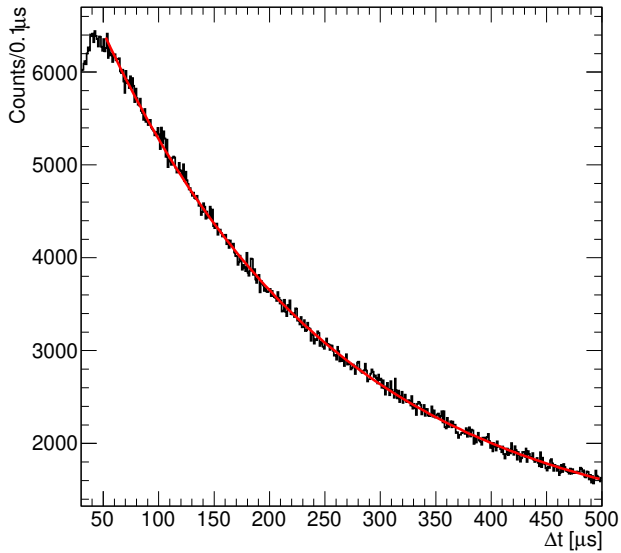}
    \caption{$\Delta t$ (black dots) and resulting fit (grey solid line) using the function from Eq.~(\ref{eq:expfit}), for neutron events detected after muons by WIT. The time constant obtained from the fit is $\tau = 211.8 \pm 1.7~\mu$s.}
    \label{fig:neutdt}
\end{figure}

 We build a high-purity neutron sample using WIT-triggered events verifying $g_t > 0.5$, $l_t < 1.5$~m, and $E_{rec} <  5$~MeV. We estimate the number of neutrons in this sample by comparing the distribution of time differences between neutron candidates and their parent muons, $\Delta t$, to the one expected from calibration studies using an americium beryllium (AmBe) source~\cite{Super-Kamiokande:2008mmn}. This $\Delta t$ distribution is shown in Fig.~\ref{fig:neutdt} and was fitted from 50~$\mu$s to 500~$\mu$s by the following equation
\begin{equation}
    N(\Delta t) = A\cdot e^{-\Delta t/\tau} + C
    \label{eq:expfit}
\end{equation}
where $N$ is the number of events, $\tau$ is the exponential decay time constant, and the constant {\it C} absorbs remaining background contributions. For the WIT data the time constant was measured to be $\tau = 211.8 \pm 1.7~\mu$s. In comparison, in AmBe calibration studies, the neutron capture time on hydrogen was measured to be $\tau = 203.7 \pm 2.8~\mu$s. This results in about a 2.5$\sigma$ difference between the two measurements. This discrepancy is believed to be due to missing neutrons within higher multiplicity showers from pile up. WIT defines a 1.5~$\mu$s window around a triggered event (500~ns before and 1,000~ns after the trigger time). If another neutron capture happens within that window, no new trigger will be issued. This introduces a form of deadtime~\footnote{We call the signal loss of the spallation cut due to accidental coincidence with a muon ``deadtime''.} for captures close together in time, biasing for a longer capture time constant.

\subsubsection{Identifying neutron clouds}
\label{sec:hadronic_spall}
After having successfully identified neutrons using WIT triggers, we will now investigate correlations between neutron clouds and spallation isotopes. Here, we define neutron clouds as groups of two or more WIT events observed within $500~\mu$s after a muon. Additionally, we require at least one of these WIT events to have $\Delta t > 20~\mu$s, $g_t > 0.5$, and $E < 5$~MeV. Note that fewer events reconstruct outside the ID (and even the fiducial volume) due to the WIT trigger conditions on the online event reconstruction.
As detailed at the beginning of this section, these cuts allow to discriminate neutrons against noise fluctuations, radioactivity, and afterpulsing events.

If a cloud was found using the conditions listed above, spallation candidates were searched for in close proximity of the center of a neutron cloud. Spallation candidate events were preselected using the noise reduction and quality cuts described in Sec.~\ref{sec:solaranalysis} and in Ref.~\cite{skivsolar} for the SK-IV solar neutrino analysis ---without the spallation and pattern likelihood cuts. Spallation candidates found within less than 5~m and 60~s \emph{after} an observed neutron cloud are selected. 
60~s was chosen to contain more than 99\% of the $^{16}$N decays and the 5~m value reflects the general size of muon-induced showers, as seen in Fig.~\ref{fig:ltvln}, 
In addition to this signal sample, we build a background sample using candidates found within 5~m of and up to 60~s \emph{before} neutron clouds. This background sample allows to estimate the fraction of spurious pairings between spallation candidates and uncorrelated neutron clouds in the signal sample, and subtract off the corresponding effects.

\begin{figure}
    \centering
    \includegraphics[width=\linewidth]{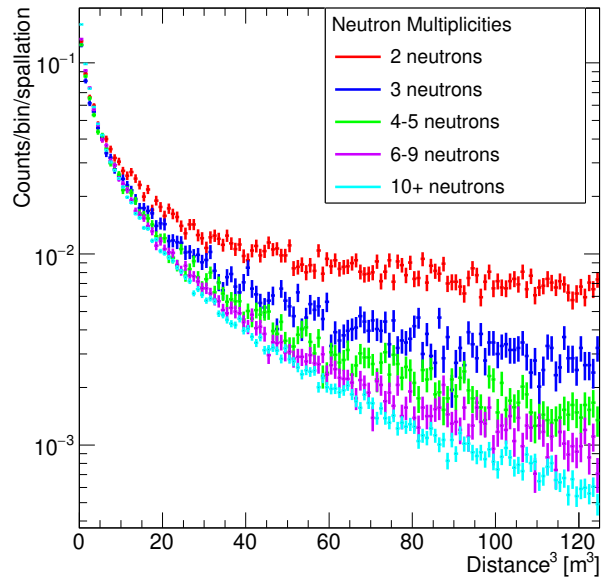}    
    \caption{Signal distribution for vertex correlation by neutron cloud multiplicity. The figure is normalized by the number of spallation events for each multiplicity (Total events in signal $-$ background distribution). As multiplicity increases, the steepness in the tail of the distribution increases as there is more accuracy in parameterization of the cloud and more spallation is expected in larger hadronic showers. Note that the vertical axis is on a logarithmic scale.}
    \label{fig:spallvertex}
\end{figure}

\begin{figure}
    \centering
    \includegraphics[width=\linewidth]{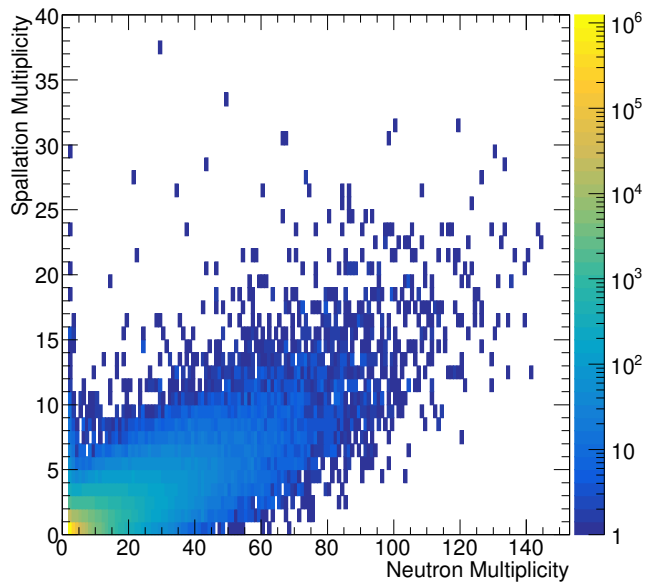}    
    \caption{Two-dimensional distribution showing the correlation between neutron capture candidate multiplicity and spallation candidate multiplicity. Note the color scale is on a logarithmic scale.}
    \label{fig:neutvspall}
\end{figure}

A strong spatial correlation between spallation candidate events and the centers of neutron clouds was found. As shown in Fig.~\ref{fig:spallvertex}, this correlation increases with the multiplicity (the number of WIT triggered events) of the neutron cloud. The slow decrease of the number of spallation candidates relative to the others at low multiplicities is due to accidental pairings between candidates and neutron clouds. Conversely, high-multiplicity neutron clouds can be more easily located and allow to reliably identify spallation products. These high-multiplicity clouds are also likely to be associated with multiple isotopes in large hadronic showers, as can be observed from Fig.~\ref{fig:neutvspall}. 

Identifying neutron clouds and correlating them with spallation candidates using the criteria outlined above allows to remove $55\%$ of spallation events with a little more than $4\%$ deadtime. Due to the low detection efficiency for the 2.2~MeV $\gamma$s resulting from neutron capture on hydrogen, small showers are missed, and therefore this procedure alone does not suffice to remove spallation in an SK analysis. In Sec.~\ref{sec:spacuts} we will show how to optimize neutron cloud reconstruction and associate it to traditional spallation cuts for the SK solar analysis. There, in addition to the observables used for this case study, we will notably use the directional goodness $g_p$ to better identify well-reconstructed neutrons, and will account for the elongated shape of neutron clouds when investigating spatial correlations with isotope vertices. 

Although the neutron cloud cuts presented here cannot be used as a standalone spallation reduction technique in SK-IV, their impact is expected to drastically increase in the new SK-Gd phase of the detector. Estimating this new impact and redesigning spallation cuts accordingly requires extensive simulation studies. In the next section, we compare neutron cloud measurements in pure water to predictions from the simulation described in Sec.~\ref{sec:simulation}, and show that this simulation can be reliably used to optimize future neutron cloud cuts.
\section{Simulated hadronic showers}
\label{sec:simucompare}
When comparing SK-IV data to the simulations described in Sec.~\ref{sec:simulation} we neglect effects associated with muon bundles and consider only single muons traveling through the whole detector, as previously mentioned. Bundles make up to $5\%$ of the total number of muons and only one of them is expected to produce a shower; therefore for moderate-size bundles the neutron multiplicity is not expected to change. In order to accurately model muon showers and reconstruction effects in the simulation we use the following two samples:
\paragraph{Hadron producing muons:} we define hadron producing muons as muons that undergo an inelastic interaction with a nucleus in water leading to the production of two or more particles of the type: $\gamma$-rays, pions, kaons, or neutrons. These particles often initiate a cascade. These showers can be observed in about $11\%$ of the muons generated using the procedure described in Sec.~\ref{sec:simulation} and $5\%$ of them lead to the producton of isotopes contributing to backgrounds in low energy SK searches. Spallation backgrounds are hence induced by about $0.5\%$ of the muons passing through SK. For this study we generate a sample of $2.7\times 10^5$ muons, $97\%$ of which are reconstructed as single through-going muons~\footnote{The remaining $3\%$ are mostly stopping muons. A few muons are also identified as bundles with only one fitted track.}.
\paragraph{Electromagnetic only muons:} this category includes muons that either undergo only minimum ionizing interactions or induce purely electromagnetic showers. Here we generate a sample of $7.5\times 10^5$ muons. Since we do not need to model hadronic showers and isotope production, we do not use FLUKA to propagate particles in water and directly combine MUSIC with the detector simulation in the pipeline described in Sec.~\ref{sec:simulation}. The fraction of single through-going muons in the final sample is only $87\%$, as muons in this sample are typically less energetic than spallation-inducing muons and more likely to stop inside the detector.

We use the analysis procedure detailed in Sec.~\ref{sec:neutspall} to extract observables characterizing the neutron cloud shapes, and the neutron multiplicity in muon-induced showers. Comparing the simulation results to the SK-IV data in pure water will allow us to estimate the systematic uncertainties associated with the neutron modeling, and motivate the use of a similar simulation to optimize spallation cuts for the current SK-Gd phase~\cite{SK-Gdloading}.

\subsection{Neutron clouds}
Using the cuts described in Sec.~\ref{sec:neutspall}, we identify neutron clouds in the SK-IV data and in the simulation. We use the simulation to evaluate the performance of the neutron tagging algorithm, then characterize the shapes of the neutron clouds as well as their multiplicities.

\subsubsection{Neutron tagging algorithm performance}
We use the MC true neutron capture times from showering muons to determine the neutron trigger efficiency and reconstruction accuracy. We consider a neutron to be correctly reconstructed if a signal is found within $50$~ns of the simulated capture time. The tagging efficiencies associated with the WIT trigger and the cuts described in Sec.~\ref{sec:softtrig} and Sec.~\ref{sec:neutspall} are shown in Table~\ref{tab:tab_efficiency}. The final neutron identification efficiency is around $6.5\%$. This low efficiency is due to the weakness of the neutron capture signal ---that is often indistinguishable from dark noise--- and to our lack of a dedicated trigger to save PMT hits following muons. The associated mistag rate can be estimated using random trigger data, and evaluating the efficiency of the $l_t$ cut using the minimum ionizing muon sample. We find a rate of $0.044 \pm 0.001$ tagged fake neutrons per EM shower muon. Conversely, the rate of real neutrons tagged per hadron-producing muon is $0.240 \pm 0.005$.

\begin{table}
	\begin{center}
	\caption{\label{tab:tab_efficiency}Neutron tagging efficiencies for different selection criteria applied sequentially: WIT trigger, proximity to the muon track, timing cut to remove the primary muon signal and the associated after-pulse, goodness-of-fit cut, and reconstructed energy ($E_n$) cut. We consider a simulated neutron to be tagged if its true capture time lies within 50~ns of the reconstructed capture time of a neutron candidate.}
\begin{tabular}{cc}
	\toprule
	Selection requirements & Efficiency values ($\%$)      \\ 
	\hline
	WIT trigger & 13.3 $\pm$ 0.1 \\ 
	20~$\mu\mathrm{s}<  \Delta t < 500~\mu\mathrm{s}$ & 89.4 $\pm$ 0.4 \\ 
	$l_t<500$~cm & 76.7 $\pm$ 0.3 \\ 
	FV  & 93.3 $\pm$ 0.5 \\ 
	Fit quality & 76.9 $\pm$ 0.4 \\ 
	$E_n <$ 5~MeV & 99.9 $\pm$ 0.6 \\ 
	\bottomrule        
\end{tabular}
\end{center}
\end{table} 
\subsubsection{Neutron cloud shapes}
We define the shape of a neutron cloud using the $l_t$ and $l_{\mbox{\tiny LONG}}$ observables defined in Sec.~\ref{sec:vardef} and shown in Fig.~\ref{fig:vardia}. For the latter, in order to compare multiple clouds, we use the average $l_{\mbox{\tiny LONG}}$ of the cloud as a reference and consider $\Delta l_{\mbox{\tiny LONG}} = l_{\mbox{\tiny LONG}} - \langle l_{\mbox{\tiny LONG}}\rangle$. In what follows we define a neutron cloud as a cluster of two or more reconstructed neutrons. In the simulation these reconstructed clouds represent only about $5\%$ of all real neutron clouds, due to the low neutron identification efficiency.

In spite of the low neutron mistag rate, due to the large fraction of minimum ionizing muons the data sample will be contaminated by non-negligible contributions from fake neutrons. Estimating the contribution from these fake neutrons requires determining the fraction of muons not leading to hadronic showers (referred to as electromagnetic-only muons at the beginning of this section), which is determined by nuclear effects that are difficult to model accurately~\footnote{For example, as will be shown in Sec.~\ref{sec:yields}, we expect $\mathcal{O}(100\%)$ uncertainties on the predicted isotope yields.}. This fraction, however, can be readily extracted from data since fake neutrons will populate the tails of the $l_t$ and $\Delta l_{\mbox{\tiny LONG}}$ distributions while contributions from real neutrons, namely neutrons produced in muon induced showers in the simulation, will dominate at small distances. At SK, with a muon rate of about 2~Hz, only a few months of data taking are needed for this estimate. 

For this study, we evaluate the fraction of EM muons by fitting the predicted $l_t$ and $\Delta l_{\mbox{\tiny LONG}}$ distributions to the SK-IV data. We perform a separate $\chi^2$ fit for each observable in order to evaluate the robustness of our model and find that the best-fit fractions of muons without hadronic showers are $96$\% ($\chi^2/NDF = 1.1$, with NDF being the number of degrees of freedom) and $97\%$ ($\chi^2/NDF = 1.6$) for $l_t$ and $\Delta l_{\mbox{\tiny LONG}}$ respectively. These values are compatible with each other, but larger than the FLUKA prediction of $89\%$. For this analysis we use a fraction of $96.5\%$ and treat the difference between the fit results for the $l_t$ and $\Delta l_{\mbox{\tiny LONG}}$ as a systematic uncertainty. Table~\ref{tab:tab_ratesmu} shows the resulting muon rates for different tagged neutron multiplicities.
\begin{table}
	\begin{center}
		\caption{\label{tab:tab_ratesmu}Table with muon rates in SK if 0 neutrons are tagged, at least 1 neutron is tagged and at least 2 neutrons are tagged (namely a neutron cloud is found). Muon rates are expressed in Hz and expressed separately for hadron producing and electromagnetic only muons, where the fraction of the second sample is extracted from $\chi^2$ as described in the main text. Only through going muons are considered.}
		\begin{tabular}{cccc}
			\toprule 
			&  & Muon rate [Hz] if:  &  \\  
			& 0 $n$ tagged & $\ge$ 1 $n$ tagged  & $\ge$ 2 $n$ tagged  \\ 
			\hline
			Hadr. muons & 0.051 & 0.0088 & 0.0023 \\ 
			EM muons & 1.48 & 0.066 & 0.0018 \\ 
			\bottomrule        
		\end{tabular}
	\end{center}
\end{table} 

\begin{tabular}{|c|c|c|c|}
 
\end{tabular} 

\begin{figure}
    \centering
    \includegraphics[width=\linewidth]{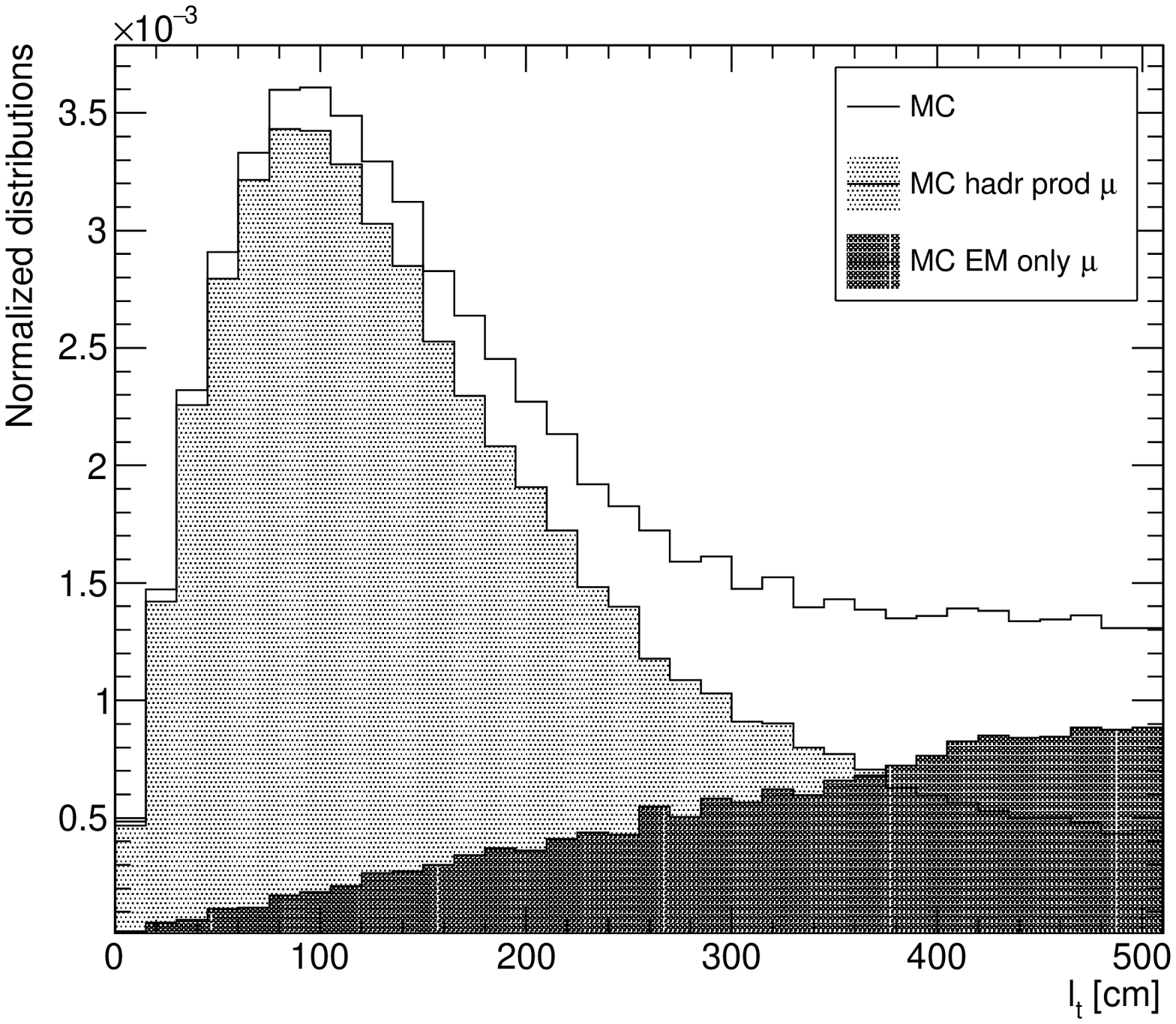}
    \includegraphics[width=\linewidth]{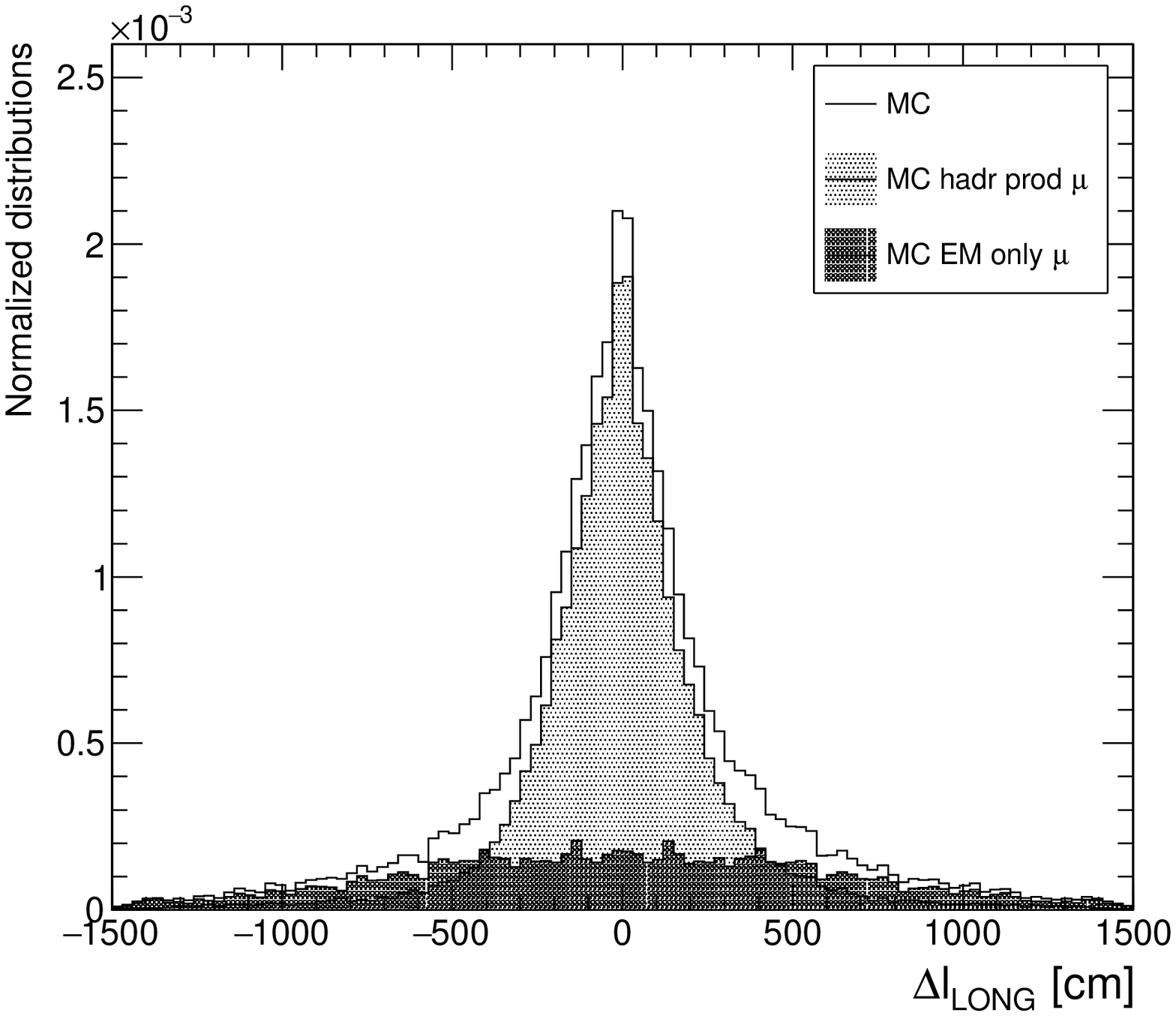}
    \caption{$l_t$ (top) and $\Delta l_{\mbox{\tiny LONG}}$ (bottom) distributions for neutrons belonging to a cluster of multiplicity larger than or equal to two, for simulation results. The fraction of muons that do not produce hadronic showers is set to $96.5\%$. The black histograms with solid line represent contributions from all reconstructed neutrons while the gray filled and black filled distributions show the true and fake neutron components respectively.}
    \label{fig:ncloudsimu}
\end{figure}

The distributions of $l_t$ and $\Delta l_{\mbox{\tiny LONG}}$ for all neutrons are shown in Fig.~\ref{fig:ncloudsimu} for the simulation. Here the uncertainties combine both the statistical uncertainties and the systematics associated with the different optimal EM muon fractions described in the previous paragraph. As expected, fake neutrons are associated with larger distances as they can be observed anywhere within $5$~m of the muon track. The real neutron component overwhelmingly dominates for $\Delta l_{\mbox{\tiny LONG}} < 5$~m and $l_t < 3$~m.

\begin{figure}[htb]
    \centering
   \includegraphics[width=\linewidth]{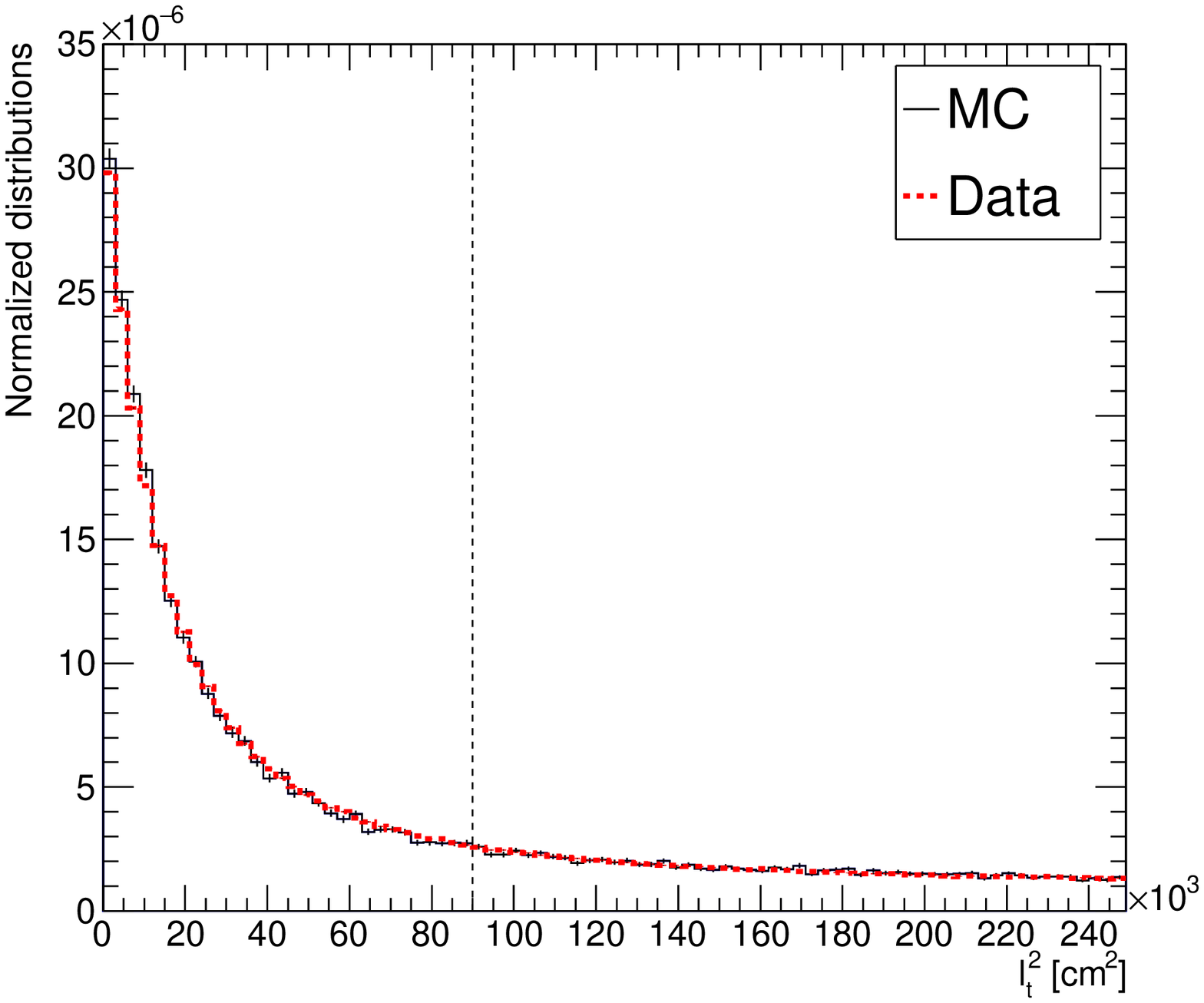}
    \includegraphics[width=\linewidth]{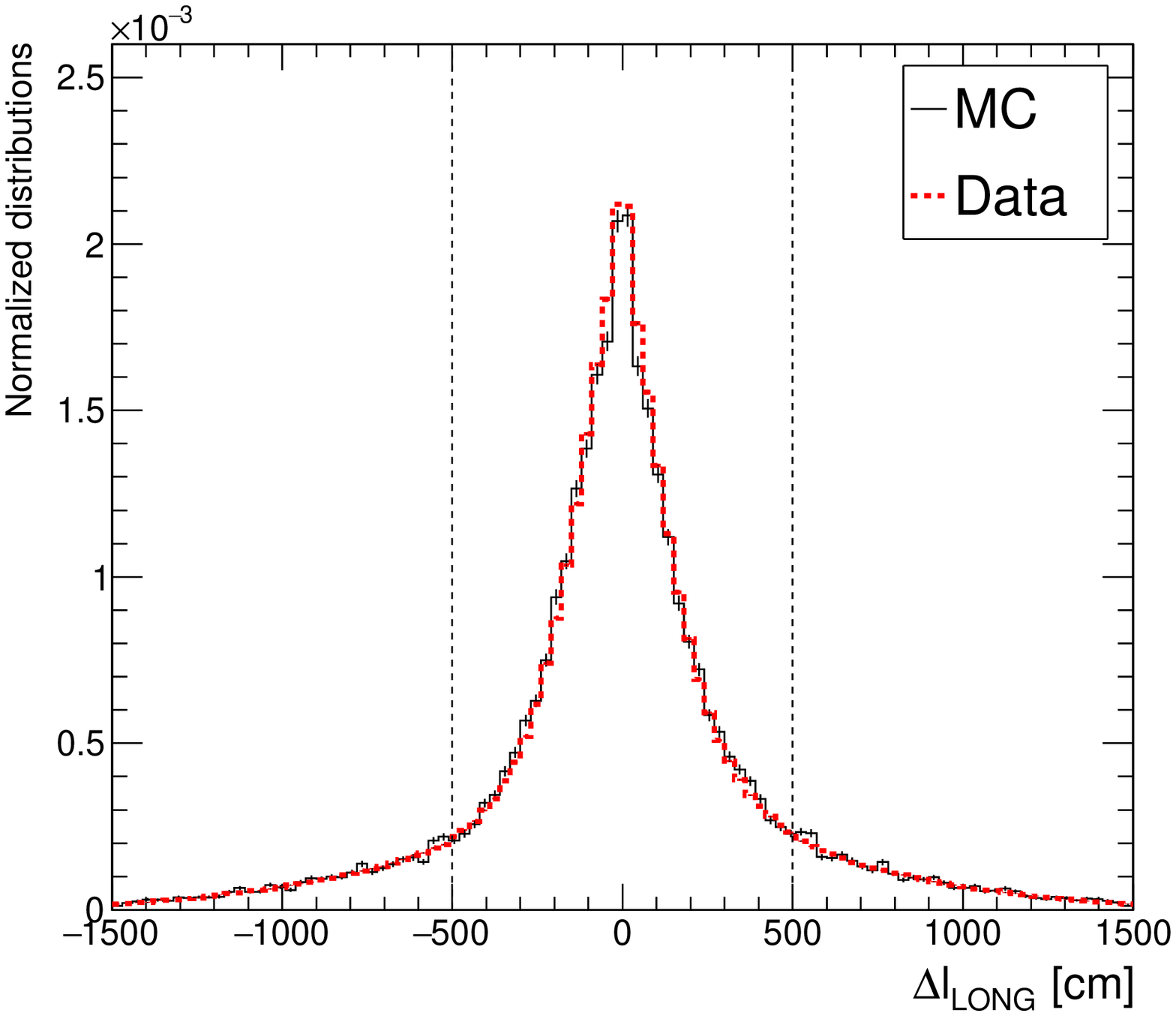}

    \caption{$l_t^2$ (top) and $\Delta l_{\mbox{\tiny LONG}}$ (bottom) distributions for neutrons belonging to a cluster of multiplicity larger than or equal to two, for simulations (black) and data (red). The fraction of muons that do not produce hadronic showers is set to $96.5\%$. The dotted lines the average transverse and longitudinal cloud extensions of $3$~m and $5$~m respectively.}
    \label{fig:neutroncloudcompare}
\end{figure}

Figure~\ref{fig:neutroncloudcompare} shows the $l_t^2$ and $\Delta l_{\mbox{\tiny LONG}}$ distributions for the SK-IV data and the simulation. The observed neutron clouds have an elongated elliptical shape, with average transverse and longitudinal extensions of $3$~m and $5$~m respectively. For distances of less than about 5~m, where contributions from real neutrons dominate, the predictions differ from the data by at most $15\%$. This excellent agreement motivates the use of a FLUKA-based simulation to predict neutron cloud shapes and optimize the associated cuts for future spallation analyses, notably at SK-Gd and Hyper-Kamiokande~\cite{HyperK}. At SK-Gd in particular, these simulation-based studies will allow to significantly reshape the spallation reduction procedure, as gadolinium doping will sizably increase the neutron identification efficiency.

\subsubsection{Neutron multiplicity}
\begin{figure}
    \centering
    \includegraphics[width=\linewidth]{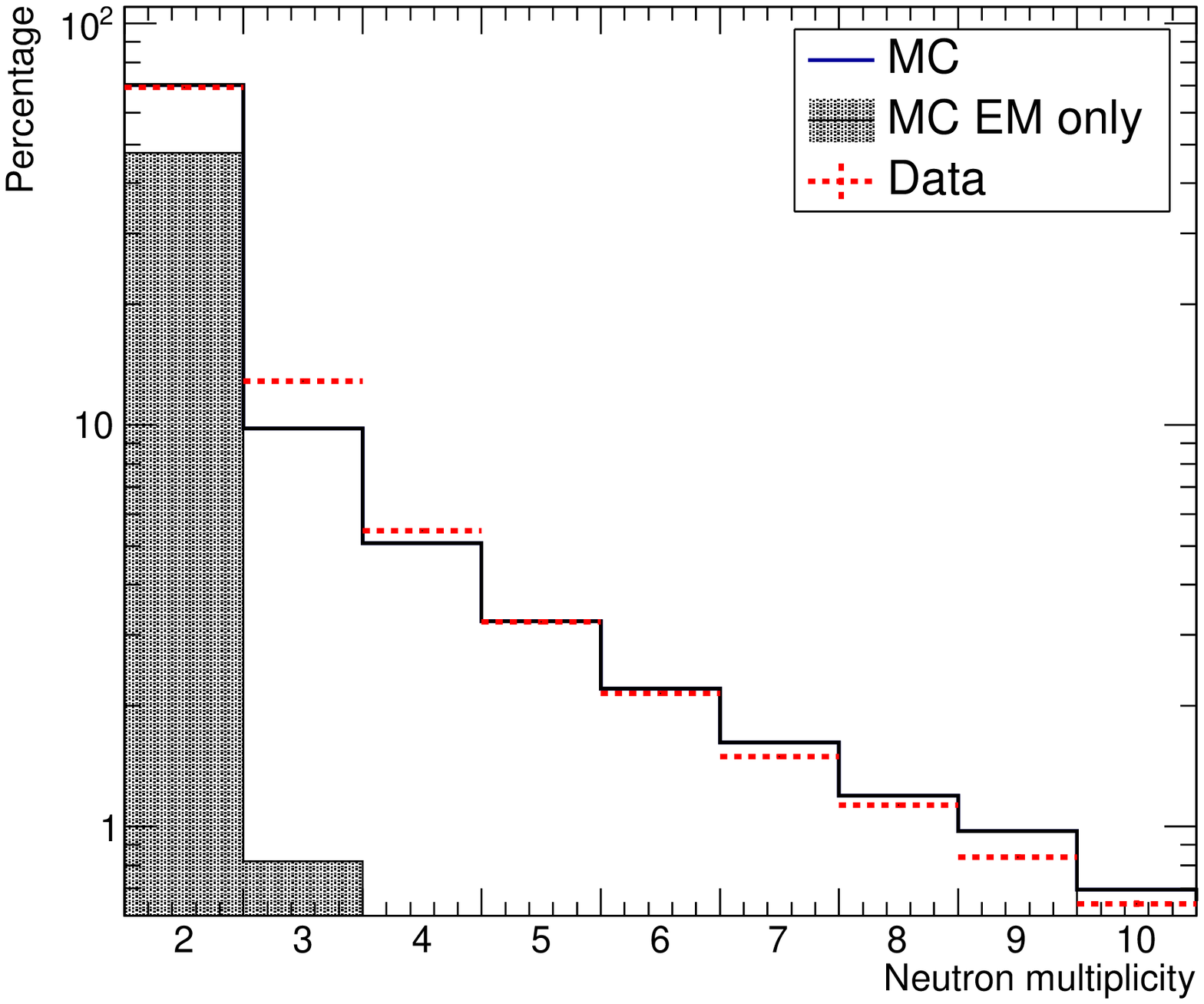}
    \includegraphics[width=\linewidth]{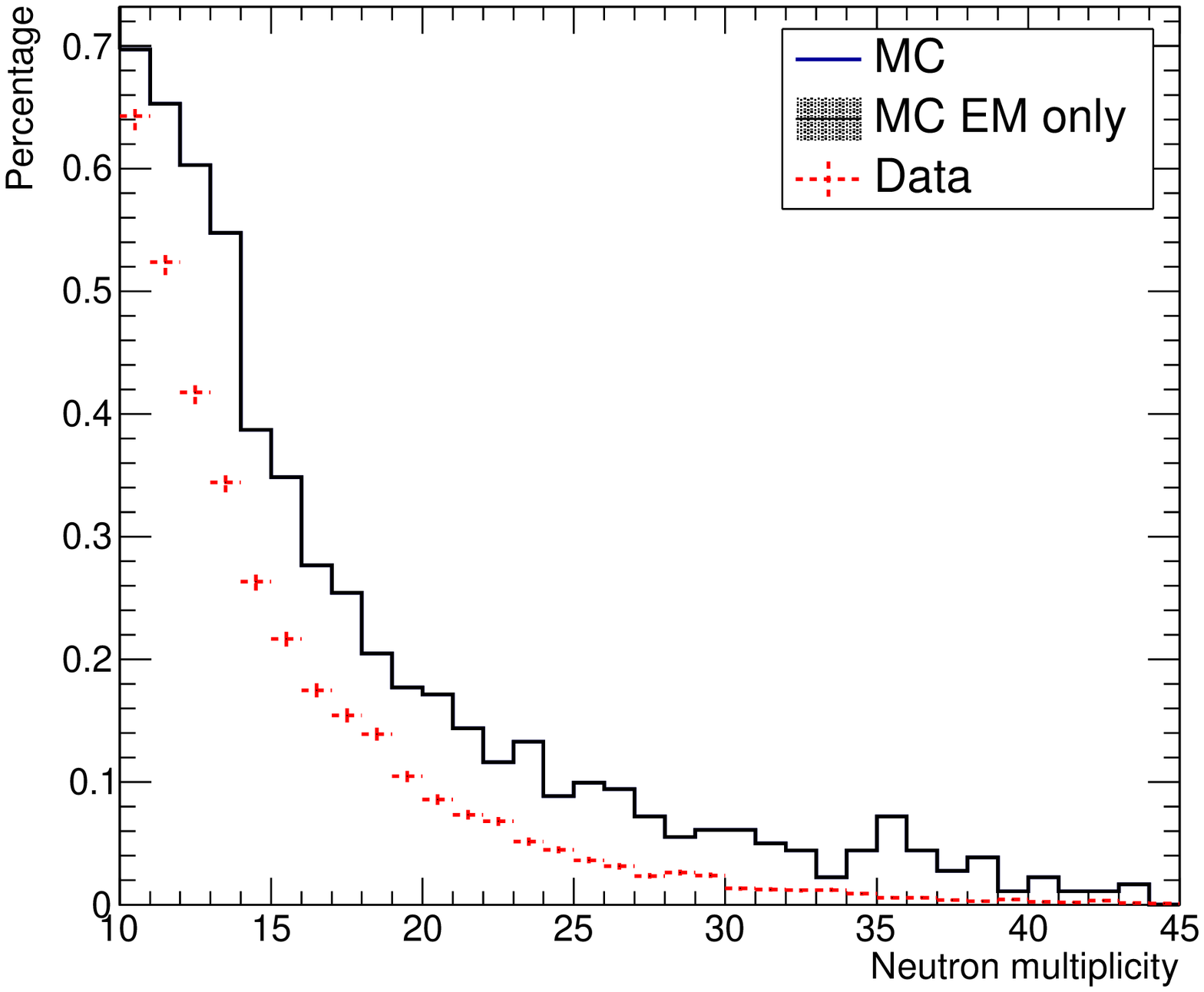}
    \caption{Neutron multiplicity distributions for simulation~(black) and data~(red) with the EM muon contribution shown for the simulation~(filled gray), this contributes only below 3 neutrons multiplicity. We show these distributions separately for neutron multiplicities from 2 up to $10$, in log scale~(top) and from 11 to $45$ in linear scale~(bottom).}
    \label{fig:neutmult}
\end{figure}

We finally estimate the number of reconstructed neutrons associated with the muons in both the data and simulation samples. Here, as in the previous section we use an EM muon fraction of $96.5\%$ and treat possible mismodeling of this fraction as a systematic uncertainty. The neutron cloud multiplicities for both simulation and data are shown in Fig.~\ref{fig:neutmult}. The abundance of low-multiplicity clouds is due to both fake neutron contributions and the low efficiency of the neutron tagging algorithm. For neutron clouds with multiplicities lower than $10$, simulation and data show reasonable agreement. 
Conversely, for multiplicities larger than $10$, FLUKA fails to accurately simulate the tails of the data distribution. Note, however, that such large clouds are typically associated with shower-producing hundreds, sometimes up to thousands of neutrons. Muons associated with these high-multiplicity showers are not only rare but also deposit a high amount of light in the detector and are hence easier to identify using other observables, such as for example the residual charge $Q_{res}$ that will be introduced in Sec.~\ref{sec:spaloglike}. In any case, neutron multiplicity distributions from data will be used to improve the simulation in future.
Our results hence demonstrate the ability of FLUKA-based simulations to accurately model hadronic showers for the types of muons that need most to be studied in future SK analyses.

\section{Solar neutrino analysis}
\label{sec:solaranalysis}
We will now use the insight gained through the studies described in this paper to design new spallation cuts for a specific SK-IV analysis. For this paper, we will focus on the solar neutrino measurements. Indeed, while many low energy analyses at SK target antineutrinos, whose interactions produce neutrons, solar neutrino interactions do not have a signature, so radioactive $\beta$ decays will look similar to them (except for the direction of the electron). Hence, even with SK-Gd, the solar neutrino analysis will continue to heavily rely on dedicated spallation tagging techniques, in addition to control of intrinsic radioactive backgrounds such as from the radon decay chain~\cite{Nakano:2019bnr}. In what follows we present an overview of the current analysis, and in particular of the previous spallation reduction strategy.

\subsection{Overview}
Here, we present a brief overview of the cuts used for the SK-IV solar neutrino analysis. A more detailed description of these cuts can be found in Ref.~\cite{skivsolar}.  

$^8$B and {\it hep} solar neutrinos from the Sun scatter elastically on electrons in SK, with the recoiling electron producing a single Cherenkov ring. The produced electrons have energies up to about 20~MeV; however, intrinsic radioactivity is a limiting background that dominates below about 5.5~MeV kinetic energy. In this paper, we therefore consider the 5.49$-$19.5 MeV energy range. The impact of these new spallation cut son the full SK-IV solar neutrino analysis will be described in~\cite{newsolarpaper}.

After high-efficiency noise reduction cuts are applied, notably rejecting events outside the FV, energy-dependent quality cuts are imposed on the reconstructed vertices and directions of the events, the goodness observables described in Sec.~\ref{sec:lowereco}. The major quality cuts select events based on their reconstruction goodness ($g_{R} = g^2_{t} - g^2_{p}$) and their \emph{effective} distance to the ID wall ---obtained by following their reconstructed direction backwards. We then check for consistency with a single electron ring (rather than, e.g. light deposited by $\beta\gamma$ decays of $^{16}$N). We parametrize the amount of multiple Coulomb scattering of the electron (the ``fuzziness" of the ring) with the ``multiple scattering goodness" to separate lower energy $\beta$ decays. Most of the remaining background events are removed using a dedicated spallation cut. 

The number of solar neutrinos can be readily extracted from the sample of events remaining after cuts by considering event directions. Indeed, electrons recoiling from neutrino elastic scattering will be almost collinear with the incoming neutrinos, that is the angle between the direction of a given event and the direction from the Sun at its detection time, $\theta_{Sun}$ is small (less than 15$^\circ$) and the $\cos\theta_{Sun}$ distribution is strongly peaked around $1$. The numbers of solar neutrinos and background events are extracted from a fit to the $\cos\theta_{Sun}$ distribution.

\subsection{Spallation cuts for the previous solar analyses}
\label{sec:spaprevious}
Spallation backgrounds can be significantly reduced by identifying space and time correlations between each isotope decay and its production in a hadronic shower initiated by a muon. To this end, each solar neutrino candidate selected using the noise reduction and quality cuts described above is paired with muons detected up to 100~s before it. For each pair, three observables are then considered: the time difference $\Delta t$ between the solar neutrino event candidate and the muon, the transverse distance of the candidate to the muon track $l_t$ ---defined in Fig.~\ref{fig:vardia}--- and the residual charge $Q_{res}$, defined as the excess charge deposited by the muon in the detector compared to the expectation from minimum ionization. For each observable, probability distribution functions~(PDFs) are then defined for spallation pairs, formed by isotope decay events and their parent muons, and for uncorrelated ``random'' pairs. These PDFs then allow to define a log likelihood function $\log_{10}\mathcal{L}$, whose functional form is as follows:

\begin{equation}
    \log_{10}\mathcal{L} = \log_{10} \left[ \prod_i \mbox{\it PDF}_{spall,i}(x_i)/\mbox{\it PDF}_{ran,i}(x_i) \right]
    \label{eq:loglike}
\end{equation}

\noindent where {\it PDF}$_{spall,i}$ and {\it PDF}$_{ran,i}$ designate the PDFs associated with a given observable $i$ for spallation and random pairs respectively. 

In the absence of a spallation simulation, the PDFs for spallation and random pairs need to be extracted from data. One sample is built by pairing solar event candidates with preceding muons (as described above). It will contain a mixture of spallation and random coincidence pairs. Using the times of events with energies much below 6~MeV, we construct a``random sample'' by generating a vertex from a uniform distribution filling the entire detector. When paired with preceding muons, this random sample estimates the random coincidence contribution, so the corresponding PDFs are extracted. Also, after subtraction of the random sample, the spallation PDFs are extracted from the data sample. Alternatively, we invert the time sequence and pair solar candidates with muons up to 100~s \emph{after} them. This inverted sample is used the same way as the random sample. Finally, in order to account for possible correlations between observables, the $l_t$ PDFs are computed for seven different $Q_{res}$ bins. Since the muon fitter used for these cuts considered only single through-going muons, a goodness-of-fit cut was also considered for this PDF in order not to be misled by poorly-fitted muon tracks. 

The final likelihood cut used for the solar analysis removed 90\% of spallation events with a position-averaged 20\% deadtime. This deadtime was measured with the random sample as a function of position. In the next section, we will show how the SK-IV new electronics, the new techniques described in this paper, and better muon reconstruction algorithms allow to further reduce this deadtime for the upcoming analysis.
\section{Spallation cuts for the solar analysis}
\label{sec:spacuts}
Here, we present a new spallation cut that improves on the reduction strategy described in Sec.~\ref{sec:spaprevious}. We take advantage of several improvements and studies that took place within the last decade. First, the muon track reconstruction was replaced. Previously we used a simple, fast muon track fitter developed at the beginning of SK. It assumes through-going single muons and misreconstructs or fails on other muons. The more complex muon track reconstruction of this analysis categorizes as described in Sec.~\ref{sec:muons}, and reconstructs all categories (up to ten tracks). It was used to reject spallation background by dE/dx reconstruction in the search for diffuse supernova neutrino interactions in SK-I, II, III~\cite{bib:sksrn123}, which
inspired the development of
FLUKA-based simulation studies~\cite{BLi_1} and highlighted the importance of muon-induced hadronic showers for isotope identification, and allowed to characterize their shapes and sizes. Finally, the improvements in the detector electronics associated with the SK-IV phase allowed to raise the PMT saturation rate and detect higher values of the total charge (and therefore $Q_{res}$), as well as identify neutron clouds as described in Sec.~\ref{sec:neutspall}.

The new spallation reduction strategy proceeds as follows. First, we apply two sets of preselection cuts in order to remove a sizable fraction of spallation events with minimal harm to the signal efficiency. These cuts aim at removing events close in space and time to neutron clouds, as well as clusters of low energy events, typically associated with the decays of multiple isotopes produced by the same muons. Then, we remove most of the remaining spallation events using an updated version of the likelihood cut described in Sec.~\ref{sec:spaprevious}. 

\subsection{Neutron Cloud Spallation Cut for Solar Analysis}
\label{sec:cloudcut}
Using the observables defined in Sec.~\ref{sec:vardef}, we define a set of cuts to reliably identify neutron clouds and investigate their space and time correlation with solar event candidates (most of which are spallation events before cuts). First, we define neutron candidates as WIT events found less than 500$~\mu$s after a muon and within 5~m of its track. The number of these candidates gives the neutron cloud multiplicity. Then, in order to compute the cloud barycenter, we consider a high-purity subsample of these neutron candidates, requiring them to verify $\Delta t >  20~\mu$s and $E_{rec} < 5$~MeV. We then assign weights to these candidates depending on their vertex and direction goodness $g_t$ and $g_p$. Specifically, we consider three regions of weights $0$, $1$, and $2$ in the $g_t-g_p$ space, that are shown in Fig.~\ref{fig:mcovaq}.

\begin{figure}
    \centering
    \includegraphics[width=9cm]{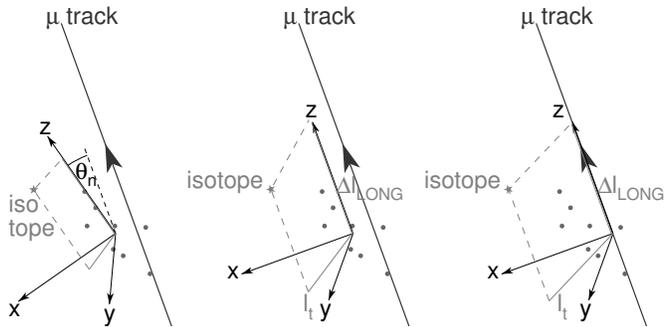} 
    \caption{Possible options for the neutron cloud coordinate systems. Left: the origin is the neutron cloud center and the axes are the neutron cloud axes. $\theta_n$ is the angle with respect to the muon track. Center: the origin is the neutron cloud center and the $z$ axis is aligned with the muon track. Right: the origin is the projection of the neutron cloud center along the muon track and the $z$ axis is aligned with the muon track. We also show the definition of the transverse and longitudinal distances $l_t$ and $\Delta l_{\mbox{\tiny LONG}}$, between a spallation isotope and the origin.}
    \label{fig:coordinate_systems}
\end{figure}

\begin{figure*}
    \centering
    \includegraphics[width=\textwidth]{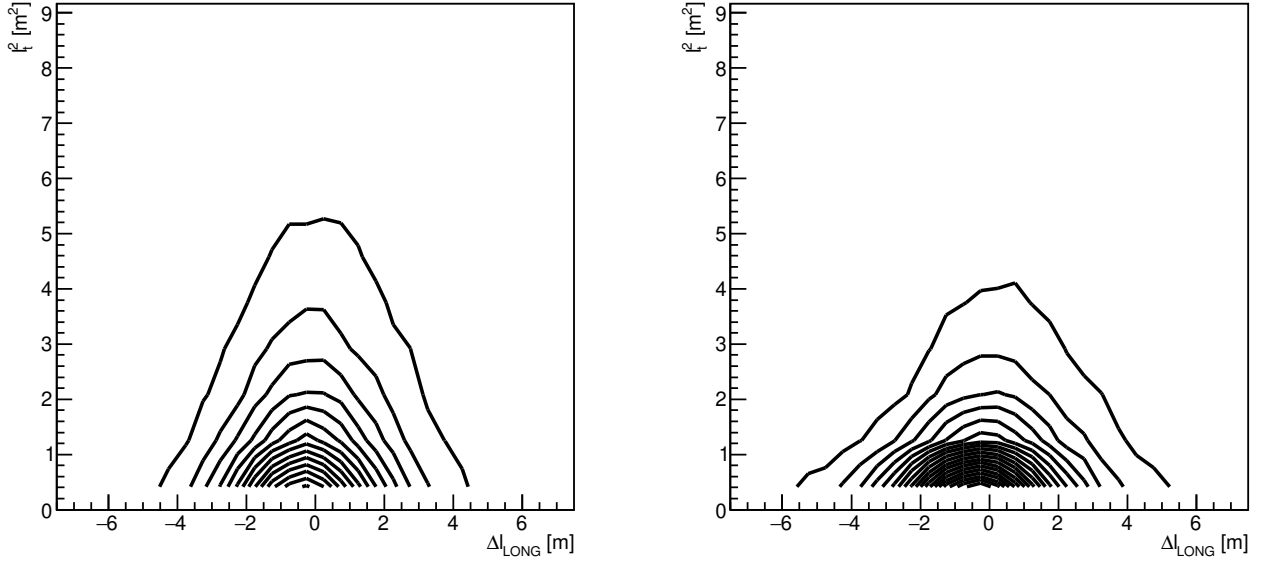}
    \caption{Tagged spallation for all neutron cloud multiplicities using a neutron cloud centered~(left) and muon track centered~(right) orientation. Greater uncertainty in lower neutron multiplicity reconstruction compared to muon track reconstruction drives the difference in ability to accurately tag spallation. Shown contour levels are arbitrary but consistent between the left and the right panel.}
    \label{fig:neutmuoncomp}
\end{figure*}

Once clouds are identified and their barycenter is defined, their positions and detection times can be compared to the location and times of solar event candidates. For this analysis we consider SK-IV low energy events that passed all the cuts defined for the solar analysis in Sec.~\ref{sec:solaranalysis} except spallation cuts. This sample is expected to be largely dominated by spallation isotope decays. We then invert the time sequence similar to the procedure described in Sec.~\ref{sec:spaprevious}, this time considering neutron clouds observed up to 60~s before each low energy event. In order to take advantage of the expected shower shape of neutron clouds, we then need to define a specific coordinate system for each cloud. Here, we consider three possible options, shown in Fig.~\ref{fig:coordinate_systems}. First, the axes of the new coordinate system could align with the axes of the best-fit ellipsoid of neutron cloud. This option is not practical, however, due to the large shape uncertainties for the low multiplicity clouds. A second possibility is to use the muon track as the $z$ axis of our coordinate system and the center of the neutron cloud as its origin. Finally, the third option also uses the muon track as a $z$ axis but sets the projection of the cloud center on the muon track as the origin. In order to assess the discriminating power of these last two options, we compute the transverse and longitudinal distances of low energy events, $l_t$ and $\Delta l_{\mbox{\tiny LONG}}=l_{\mbox{\tiny LONG}}^{\mbox{\tiny isotope}}-l_{\mbox{\tiny LONG}}^{\mbox{\tiny n-cloud}}$, defined in Fig.~\ref{fig:coordinate_systems}, to the origins of their respective coordinate systems. The distribution of $l_t$ and $\Delta l_{\mbox{\tiny LONG}}$ is shown in Fig.~\ref{fig:neutmuoncomp} for all neutron cloud multiplicities. We notice that using the projection of the neutron cloud center on the muon track as an origin significantly reduces the spread of this distribution in $l_t$, a spread that is primarily driven by contributions from low multiplicity clouds. We hence choose this definition of the origin and set the $z$ axis to be along the muon track for our analysis. 

Using the coordinate system defined on the right panel of Fig.~\ref{fig:coordinate_systems} we then define cuts on $\Delta t$, $l_t$ and $\Delta l_{\mbox{\tiny LONG}}$ for each low energy event--neutron cloud pair, with $\Delta t$ defined as the time difference between the low energy event and the muon associated with the cloud. We first define spherical cuts, removing events within either $0.2$~s and $7.5$~m or $2$~s and $5$~m of clouds with 2 or more neutrons. Then, we define multiplicity bins of 2, 3, 4--5, 6--9, and $\geq$10~neutrons candidates, and, for each bin, define a specific ellipsoidal cut on {\it lt} and {\it ln}. Since clouds with only 2~candidates are often not associated with hadronic showers, as shown in Sec.~\ref{sec:simucompare}, we require $\Delta t < 30$~s. For higher multiplicities we consider all muons up to 60~s before the low energy event. The different cuts are summarized in Table~\ref{tab:cloudtab}. If a low energy event--neutron cloud pair correlation is within the required $\Delta t$, $l_t$ and $l_{\mbox{tiny LONG}}$ values, the corresponding low energy event is rejected.

\begin{table}[t]
    \centering
        \caption{Table showing the different cut conditions for the cloud cut. The symbol `+' is used to represent showers of at least that multiplicity.}
    \begin{tabular}{ c*{6}{c}c }
    \toprule Multiplicity  & 2+ & 2+ & 2 & 3 & 4--5 & 6--9 & 10+ \\\hline
    \% of Showers & 100 & 100 & 73 & 15 & 6 & 3 & 3  \\\hline
    $\Delta t$  & 0.2 & 2 & 30 & 60 & 60 & 60 & 60 \\\hline
    $\Delta l_{\mbox{\tiny LONG}}$ [cm] & 750 & 500 & 350 & 500 & 550 & 650 & 700  \\\hline
    $l_t$ [cm] & 750 & 500 & 200 & 245 & 346 & 447 & 500 \\\bottomrule        
    \end{tabular}
    \label{tab:cloudtab}
\end{table}

Since the sample used for this analysis is largely dominated by spallation isotope decays, the background rejection rate of the neutron cloud cut can be readily estimated. To evaluate the deadtime, we use a sample of events with reconstructed energies between 3.5~MeV and 5~MeV whose vertices have been replaced by randomly generated ones. We then pair these events with muons observed up to 60~s after them and apply the reduction steps outlined above, with the sign of $\Delta t$ inverted. The deadtime is then given by the fraction of remaining low energy events and has been found to 1.3\%.

\subsection{Multiple Spallation}
\label{sec:multispa}
\begin{figure}
        \includegraphics[width=\linewidth]{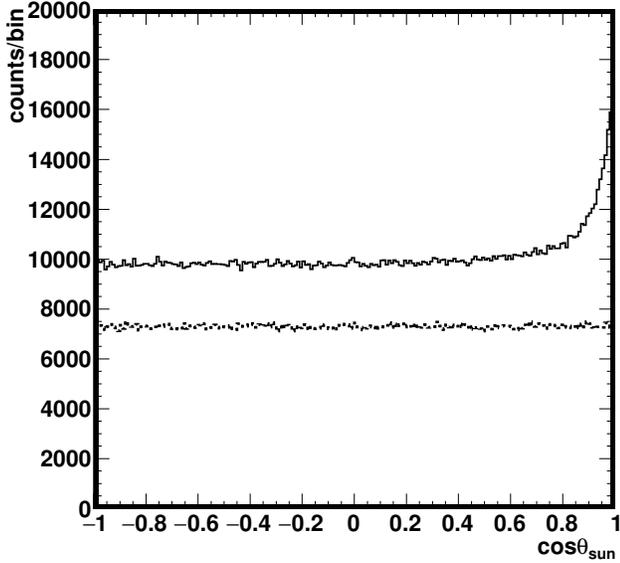} 
    \caption{Comparison of the events removed~(dashed) and remaining~(solid) in SK-IV solar sample using multiple spallation cut above 5.99~MeV . The sample above uses the final sample criteria from \cite{skivsolar}.}
    \label{fig:multispa}
\end{figure}

Since neutrons in a hadronic shower are a good indicator of
the production of spallation nuclei, then the observation of the decay of a spallation nucleus is likewise an indicator for a hadronic shower. We therefore apply
(in addition to the neutron cloud cut), a preselection removing clusters of isotopes produced by the same muon. Instead of pairing candidate events with possible parent muons, this multiple spallation cut identifies clusters of low energy events observed within a few tens of seconds and a few meters of each other. Here, we consider a sample composed of all SK-IV events passing the first reduction and quality cuts defined for the solar analysis, as discussed in Sec.~\ref{sec:solaranalysis}. Since we need to take all spallation isotope decays into account we apply neither the old spallation cut nor the pattern likelihood cut, as the latter targets $\mathrm{^{16}N}$ $\beta\gamma$ decays. We find that the optimal cut removes candidates found within $4$~m and $60$~s of any event from this sample. This cut allows to remove $45\%$ of spallation background events with a deadtime of $1.3\%$. The solar angle distribution of the rejected and the remaining events is shown in Fig.\ref{fig:multispa}. The absence of a peak around $\cos\theta_{sun} = 1$ for the rejected events confirms the low deadtime for this cut.  

\subsection{Spallation Log Likelihood}
\label{sec:spaloglike}
Multiple spallation in tandem with tagging neutrons from hadronic showers allow to remove 65\% of spallation events with 2.4\% deadtime. To identify the remaining spallation background events, we update the likelihood cut defined in Sec.~\ref{sec:spaprevious}. In particular, in addition to the $dt$, $l_t$, and $Q_{\mbox{\tiny res}}$ observables, we also consider the difference in longidudinal distance $\Delta l_{\mbox{\tiny LONG}}$ between an isotope and the segment of the muon track associated with the highest amount of light deposited in the detector. This new observable allows to estimate the distance between isotopes and muon-induced showers. In what follows, we describe how we build and parameterize PDFs using low energy events passing the multiple spallation and neutron cloud cuts described above. Similarly to the procedure described in Sec.~\ref{sec:spaprevious} we pair these events with muons within $60$~s~(inverting the time sequence to separate the random coincidence component from the spallation component) to obtain the PDFs distributions for spallation and random pairs. The choices of parameters of the PDFs for all observables are shown in appendix~\ref{appendix_fitparam}.

\subsubsection{Time difference $\Delta t$}
We obtain the $\Delta t$ PDFs following a procedure similar to the one described in Sec.~\ref{sec:yields}, fitting the $\Delta t$ distribution for low energy events found within 2~m of a muon track. The $\Delta t$ distribution for uncorrelated random pairs is flat in this instance is subtracted statistically. The functional form for the spallation PDF is as follows:
\begin{equation}
    \mbox{\it PDF}_{sig}(\Delta t) = \sum_i^7 A_i e^{-\Delta t/\tau_i}
\end{equation}
where $\tau_i$ is the decay constant for the isotope and $A_i$ is the fitted amplitude. 

\subsubsection{Transverse distance $l_t$}
\begin{figure*}
    \centering
    \includegraphics[width=\textwidth]{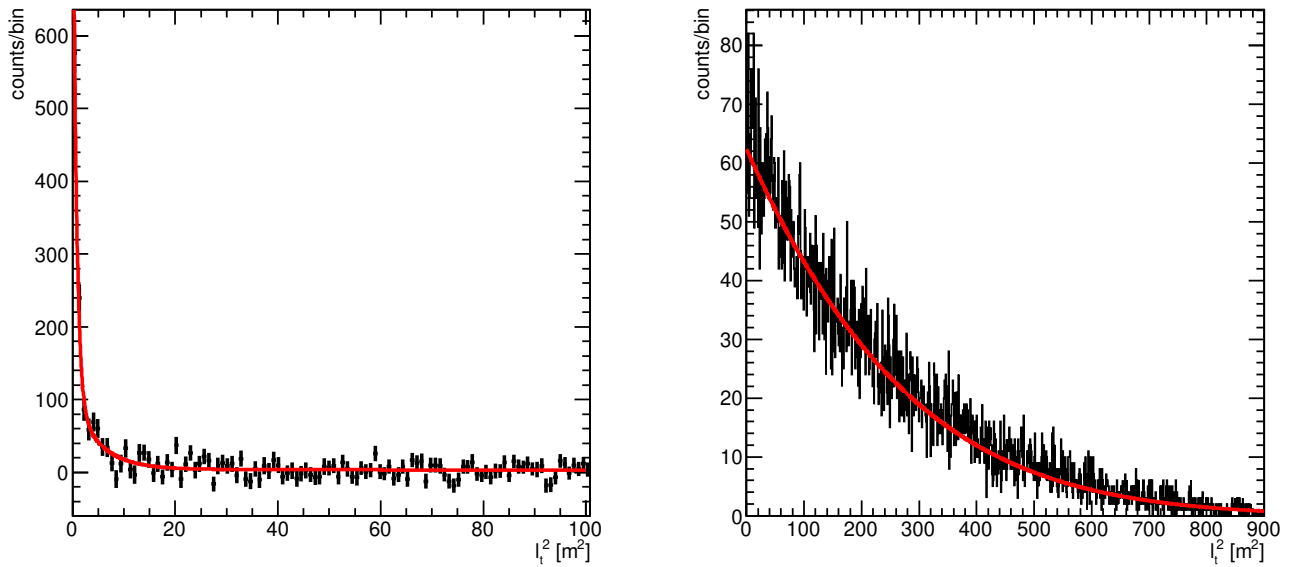}
    \caption{Distributions with PDF for the spallation (left, normal time sequence) and random coincidence~(right, inverted time sequence) samples with corresponding fits (solid lines). The distributions shown here are for the 3--30~s $\Delta t$ and 0.5--1.0~Mpe $Q_{\mbox{\tiny res}}$ bin. The analytical forms for the fits are shown in Equations~\ref{fitltsig} and~\ref{fitltbg}.}
    \label{fig:lt2both}
\end{figure*}

To account for correlations between $l_t$ and other observables we compute PDFs for $l_t$ in 7 $Q_{\mbox{\tiny res}}$ and 3 $\Delta t$ bins. These $\Delta t$ bins are taken to be 0--100~ms, 100~ms--3~s, and 3--60~s, in order to account for the different half-lives of the isotopes. To reflect the amount of phase space available, we express PDFs as a function of $l_t^2$. Their functional forms are as follows:
\begin{gather}
\label{fitltsig}
    PDF_{SIG,l_t}(l_t^2) = \sum_{i=1}^3e^{c_i - p_i\cdot l_t^2}\\
    \label{fitltbg}
    PDF_{BG,l_t}(l_t^2) = 
    \begin{cases}
       p_0 & l_t^2 \leq l_{t0}^2 \\ 
       p_0e^{-p_1(l_t^2 - l_{t0}^2) + p_2(l_t^2-l_{t0}^2)^2}          & l_t^2 > l_{t0}^2 
    \end{cases}
\end{gather}

\noindent where $c_i$ and $p_i$ are the fit parameters. Due to the finite size of the detector, at large $l_t$ the allowed region is no longer cylindrical, so a ``piecewise'' function was defined for the background PDFs, with $l_{t0}^2$ being the point where the function changes. As an example, the $l_t$ distributions and PDFs are shown for the 3--30~s $\Delta t$ and 0.5--1.0~Mpe $Q_{\mbox{\tiny res}}$ bin in Fig.~\ref{fig:lt2both}, for the spallation and random coincidence samples.

\subsubsection{Residual charge $Q_{\mbox{\tiny res}}$}
\begin{figure}
    \centering
    \includegraphics[width=\linewidth]{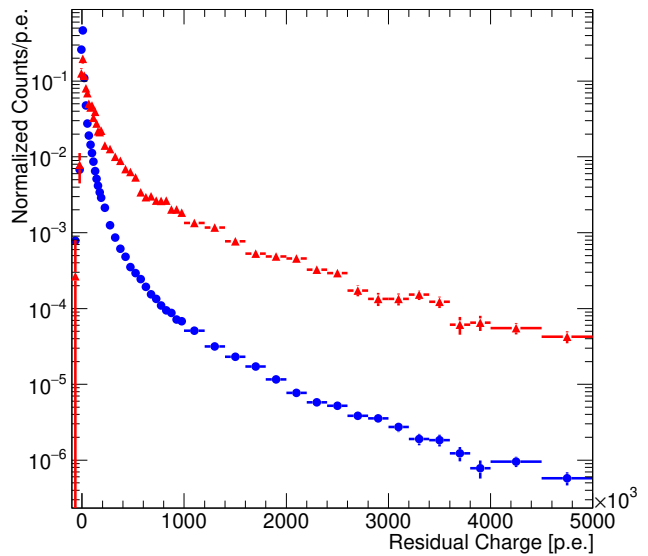}
    \caption{Signal (red triangles) and Background (blue circles) distributions for the sample used for the $Q_{\mbox{\tiny res}}$ PDF construction.}
    \label{fig:qres}
\end{figure}

The residual charge ($Q_{\mbox{\tiny res}}$) is the excess charge observed for a muon event compared to a minimum ionizing particle (MIP) traveling the same distance inside the detector. The MIP muon charge per unit of track length is defined as the peak value of the distribution of the amount of charge deposited by unit track length for single through-going muons. It is evaluated for each run time period and typically lies around $26.78$~photo-electrons~(p.e.) per cm. We obtain the PDFs for this observable by using a sample of low energy events within 2~m and 10~s of a muon. The distributions for spallation and random uncorrelated pairs are shown in Fig.~\ref{fig:qres}. Both signal and background PDFs in the positive $Q_{\mbox{\tiny res}}$ region are a sum of exponential functions:
\begin{equation}
\mbox{\it PDF}(Q_{\mbox{\tiny res}}) =\sum^5_{i=1} e^{c_i - p_i\cdot Q_{\mbox{\tiny res}}}, \ \ Q_{\mbox{\tiny res}} > 0
\end{equation}
where $c_i$ and $p_i$ are the fit parameters for exponential functions. For negative $Q_{\mbox{\tiny res}}$, no analytical form was assumed and a linear interpolation of the sample bins was used. 

\subsubsection{Longitudinal distance $\Delta l_{\mbox{\tiny LONG}}$}
Defining PDFs for the difference in longitudinal length $\Delta l_{\mbox{\tiny LONG}}=l_{\mbox{\tiny LONG}}^{\mbox{\tiny isotope}}-l_{\mbox{\tiny LONG}}^{\mbox{\tiny dE/dx peak}}$ allows to include information about the muon-induced hadronic shower into the likelihood cut. Here, we developed a new method to reconstruct the energy loss of the muon along its track (dE/dx) based on a previously published method~\cite{bib:sksrn123}. Defects and possible improvements to this method were suggested in~Ref.\cite{BLi_3}. The method presented here remedies those defects, although the improvements differ from those suggested by~Ref.~\cite{BLi_3}. Using the entry time of the muon in the detector and the PMT hit pattern, dE/dx is estimated by identifying the points of the muon track verifying the time correlation equation for each PMT hit:
\begin{equation}
t_{PMT} - t_{entry} = l\cdot c_{vac} + d \cdot c_{water} 
\label{eq:dedxsimp}
\end{equation}
where $t_{PMT}$ and $t_{entry}$ are the PMT time and muon entry time respectively, $l$ is the distance from the muon entry point to the point along the track where the light is emitted from, $d$ is the distance from the emission point to the PMT, and $c_{vac}$ and $c_{water}$ are the speeds of light in vacuum and water respectively. The dE/dx is computed for 50~cm segments of the muon track, corresponding roughly to the vertex resolution for events in the energy region of 3.49--19.5~MeV in SK. The simplest approach to estimating dE/dx is to add the charge of each PMT to each bin containing a solution of Eq.~(\ref{eq:dedxsimp}). Here, using the method proposed in~\cite{BLi_1}, we spread the charge from each hit across multiple bins to account for the PMT timing resolution. More specifically, we take the contribution, $g_{ij}$, of the $i^{th}$ PMT to the $j^{th}$ bin to be:
\begin{equation}
g_{ij} = Q_i \cdot \frac{e^{d(l_j + 25)/\lambda}}{S(\theta_{ij},\phi_{ij})} \cdot \frac{f_{ij}}{\sum_k f_{ik}}
\end{equation}
where
\begin{equation}
f_{ij} = \left|\mbox{Erf}\left(\frac{\tau(l_j) - t_i}{\sqrt{2}\sigma_i}\right)- \mbox{Erf}\left(\frac{\tau(l_j+50) - t_i}{\sqrt{2}\sigma_i}\right)\right|, 
\end{equation}
$\sigma_i$ is the timing resolution for the observed charge by the $i^{th}$ PMT, $\tau(l_j)$ is the $t_{PMT}$ that solves Eq.~(\ref{eq:dedxsimp}). For the $j^{th}$ bin boundary's $d$ and $l$, Erf is the standard error function, $Q_j$ is the charge observed by the PMT, the exponential function is water attenuation correction, $S$ is the photocathode coverage correction, and the integral of the sum is normalized to one. This procedure ensures that the charge of a given PMT hit is only counted once. 
Although error functions are used for the integral, since $\tau_j(l)$ is non-linear, the integral is not easily normalized. Care also has to be taken for the shape of $\tau_j(l)$ as it is not monotonic, therefore special cases are implemented to handle scenarios where $\tau(l) - t_i = 0$ and when d$\tau$/dl = 0.

\begin{figure}
    \centering
    \includegraphics[width=\linewidth]{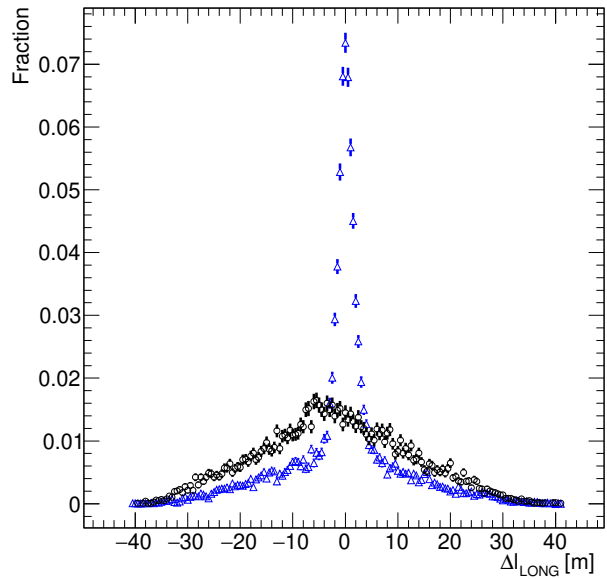}
    \caption{$\Delta l_{\mbox{\tiny LONG}}$ distribution used for PDF fit, for the spallation signal (blue triangles) and for spallation accidentals (black circles). The peak of the background distribution is shifted to negative $\Delta l_{\mbox{\tiny LONG}}$ as a result of non-showering muons being more likely to have a dE/dx peak later in the track.}
    \label{fig:ln}
\end{figure}

For muons that induce particle showers in addition to minimum ionization, the segment of the muon track associated with the largest dE/dx can indicate the location of these showers. We hence define $\Delta l_{\mbox{\tiny LONG}}=l_{\mbox{\tiny LONG}}^{\mbox{\tiny isotope}}-l_{\mbox{\tiny LONG}}^{\mbox{\tiny dE/dx peak}}$ as the longitudinal distance of an isotope to this segment along the muon track. Distributions of this observable for spallation and uncorrelated pairs are shown in Fig.~\ref{fig:ln}. To define the corresponding PDFs, the $\Delta l_{\mbox{\tiny LONG}}$ distribution for low energy events found within $2$~m and $10$~s of a muon is fitted by a sum of three Gaussians:
\begin{equation}
\mbox{\it PDF}(\Delta l_{\mbox{\tiny long}}) = \sum^3_{i=1} A_ie^{\frac{-(l_{\mbox{\tiny long}} - x_i)^2}{2\sigma_i^2}}
\end{equation}
where $A_i$, $x$, and $\sigma_i$ are the fit parameters for each PDF. For uncorrelated pairs, one of the Gaussian fits was degenerate and dropped from the final form. For minimum ionizing muons, the dE/dx peak is more likely to be at the end of the track, resulting in the background distribution being slightly shifted away from 0.

\begin{figure}
    \centering
    \includegraphics[width=\linewidth]{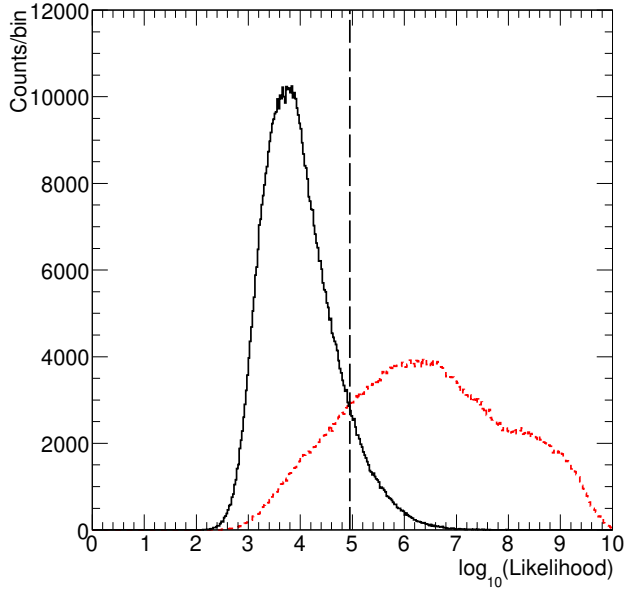}
    \caption{Comparison of the $\log\mathcal{L}$ for the signal and background of the non-neutron data period. This figure shows the distributions for the spallation signal (dashed) and spallation accidentals (solid). The vertical dashed line shows the tuned cut value. Since the multiple spallation cut is already applied, only 82\% of remaining spallation is removed to achieve 90\% overall spallation removal effectiveness.}
    \label{fig:loglike}
\end{figure}

Using the PDFs defined above, we define a log likelihood function as shown in Eq.~(\ref{eq:loglike}). The distributions of this log likelihood for signal and background are shown in Fig.~\ref{fig:loglike}. To estimate the impact of the cut on this function, we take advantage of the fact that the low energy event sample that we are considering is dominated by spallation and solar events. We can hence readily estimate the background rejection rate of our algorithm by computing the fraction of events with $\cos\theta_{sun}<0$ removed by the likelihood cuts. Conversely, the signal efficiency can be computed by applying cuts on a random sample, where low energy events are paired with muons observed after them. We use these techniques to tune the cut point for the log likelihood, maximizing the signal efficiency for a background rejection rate of 90\%, similar to the one obtained with the previous spallation cut described in Sec.~\ref{sec:spaprevious}. Since WIT was running only during a small fraction of the SK-IV period, we apply different likelihood cuts depending on whether the neutron cloud information is available. The availability of neutron cloud information notably allows to loosen the likelihood cut. 

\subsection{Total Spallation Cut Results}
After tuning the neutron cloud, multiple spallation, and likelihood cuts described throughout this section, we apply spallation reduction to a sample of SK-IV events passing the noise reduction, quality cuts, and pattern likelihood cut described in Sec.~\ref{sec:solaranalysis}.

\begin{figure}
    \centering
    \includegraphics[width=\linewidth]{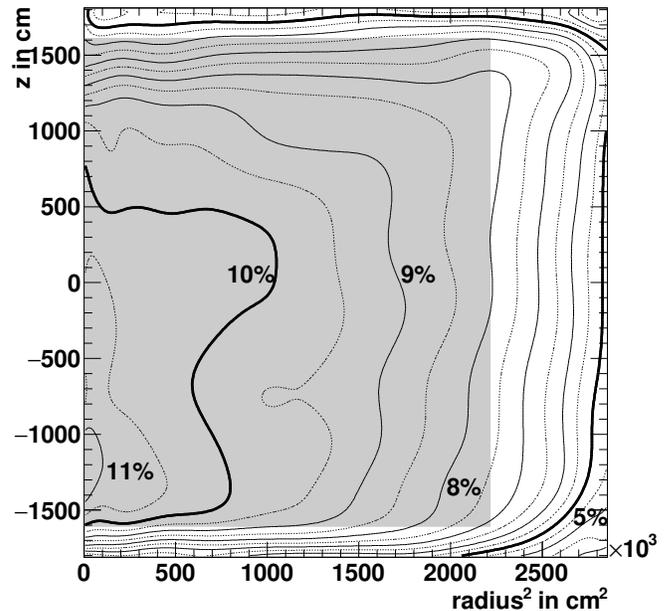} 
    \caption{Spallation cut dead time as a function of radius and height. The origin is taken at the center of the detector~\cite{skdetector}. This function combines neutron cloud, multiple spallation and likelihood dead time. The fiducial volume is indicated by the shaded area.}
    \label{fig:deadvol}
\end{figure}

The total dead times for the periods with and without neutron cloud information are 8.9\% and 10.8\%. Figure~\ref{fig:deadvol} displays the position dependence of the dead time with neutron cloud information. The new spallation cut hence allows to reduce the dead time by up to 55\% compared to the previous analysis.

\begin{figure}
    \centering
    \includegraphics[width=\linewidth]{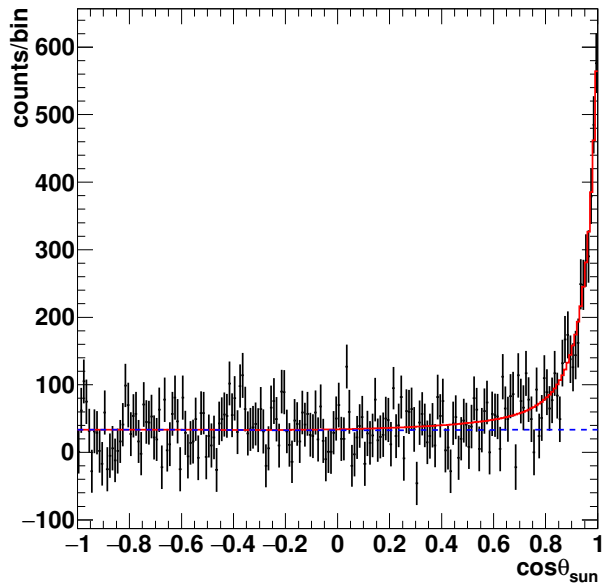}
    \caption{Figure showing the difference in events between final solar samples using new spallation cut compared to the previous cut for 5.99 to 19.5 MeV. The peak at the right contains the additional solar neutrino interactions gained with the change in cut. The small excess in the flat region reflects changes to the other solar neutrino cuts and differing interaction with the previous and new spallation cut. The intrinsic radioactivity in the energy region below 6 MeV was rejected by the previous spallation cut due to accidental coincidence.}
    \label{fig:coscompdiff}
\end{figure}

The effect of this new reduction on the SK-IV solar analysis is shown in Fig.~\ref{fig:coscompdiff}, that shows the $\cos\theta_{sun}$ distribution for events passing the new spallation cut and failing the one described in Sec.~\ref{sec:solaranalysis}. The clear peak around $\cos\theta_{sun} = 1$ shows that the new procedure allows to retrieve a sizable number of solar events. In the final sample, this cut allows for an increase of 12.6\% solar neutrino signal events, with a reduction in the relative error on the number of solar events of 6.6\%. Compared to the total SK-IV exposure, retrieving this signal corresponds to an increase of roughly a year of detector running.

\section{Isotopes Yields}
\label{sec:yields}
In this section, we estimate the spallation isotope yields at SK and compare them with the MC simulation from Sec.~\ref{sec:simulation}. To this end, we update a previous study performed using 1890 days of SK-IV data~\cite{SKspall_zhang}, which determined the yield of each isotope using their different half-lives, by fitting the distribution of the $\Delta t$ observable defined in Sec.~\ref{sec:spaloglike}. In addition to an increased livetime of $2970$ days, we update the selection criteria for the spallation event samples, as well as the computation of the cut efficiencies and isotope decay spectra.

\subsection{Selection of spallation isotopes}
\label{sec:preselyield}
To build a spallation-rich sample, we select events passing the noise reduction steps outlined in~\cite{newsolarpaper} and apply the following quality cuts:
\begin{itemize} 
    \item Events with reconstructed vertex more than 200~cm from the ID wall~(FV cut)
    \item Events less than 50~$\mu$s of the muon were rejected to remove cosmic $\mu$-$e$ decays as well as PMT afterpulsing. 
\end{itemize}

In addition to the noise reduction cuts, we apply cuts designed to increase the fraction of spallation events in the sample. First, we require the event kinetic energies to lie between 6 and 24.5~MeV. This window ranges from the energy at which spallation starts dominating over intrinsic radioactivity to the highest possible energy of the isotope decay products.

To fit the time difference between pairs of spallation isotopes and their parent muons, we pair spallation candidates with muons observed up to 100~s before them. Here, we extended the $\Delta t$ range compared to the solar analysis from Sec.~\ref{sec:solaranalysis} in order to include all long-lived isotopes. Additionally, we consider only muons with a reconstructed track of at least 200~cm. For stopping muons, we compute $l_t$ by extending the track through the entire detector; for multiple muons we considered only the primary track. Due to the 200~cm requirement, corner clipping muons will not be taken into account. Their contribution to the spallation background in the fiducial volume is however negligible; the total number of events found within 500~cm of a corner clipping muon track is about $10^{-5}$ of the total number of spallation events.

After pairing spallation candidates with suitable muons as described above, we build two separate samples with a high fraction of spallation pairs. In the first sample, we apply the same selection cut as in the previous study by Ref.~\cite{SKspall_zhang}, requiring $lt<200$~cm. In the second sample, we select muons associated with three or more tagged neutrons. In Sec.~\ref{sec:hadronic_data} events were required to have at least two events triggered in WIT within 500~$\mu$s and 500~cm, and one quality event. For this study, we instead require at least three events triggered by WIT to further increase purity. The cloud location is still only defined by the weighted center of quality neutron cloud events. Then, we pair the selected neutron clouds with spallation candidates verifying the conditions listed in table~\ref{tab:cloudtab}. Within this table, all cuts requiring $\Delta t < 60$~s were used for pairs with $\Delta t$ up to 100~s.

\subsection{Spallation Yield Fit}
\label{sec:yieldfit}
To extract isotope yields from data, we fit the decay time distributions of the 10 most abundant isotopes found in~\cite{SKspall_zhang,BLi_1} to time difference $\Delta t$ distribution using the spallation sample described above. Specifically, we parameterized the time dependence of the total event rate $R_{TOT}$ by a sum of exponentials as follows:

\begin{equation}
    R_{TOT} = \sum_i^n R_i\cdot e^{(-\Delta t/\tau_i)} + const.
\label{eq:allfit}
\end{equation}
where we keep the isotope decay constants $\tau_i$ fixed and fit the production rates $R_i$. For this study, we perform a $\chi^2$ fit analytically over the whole $\Delta t = $ 0--100~s range.  

To mitigate possible degeneracies, we grouped isotopes together when their decay constants were within $10\%$ of each other. For this study, $^8$Li and $^8$B as well as $^9$C and $^8$He were paired together. The decay constant associated with each of these pairs was taken as a weighted average between the two isotopes, using the yields predicted by~\cite{BLi_1}.
The $\tau_i$s used were 1.18~s and 0.181~s for the two pairs respectively. For the case of stopping muons, looking at the individual fit contributions, most spallation products were from those which is normally produced from a neutron interacting with nucleus. The fits are dominated by $^{16}$N, with the next largest contribution being $^{12}$B with an observed raw rate roughly a factor of six smaller. For single through going muons, $^{16}$N is observed roughly twice as often. 

\begin{figure*}
    \centering
    \includegraphics[width=\textwidth]{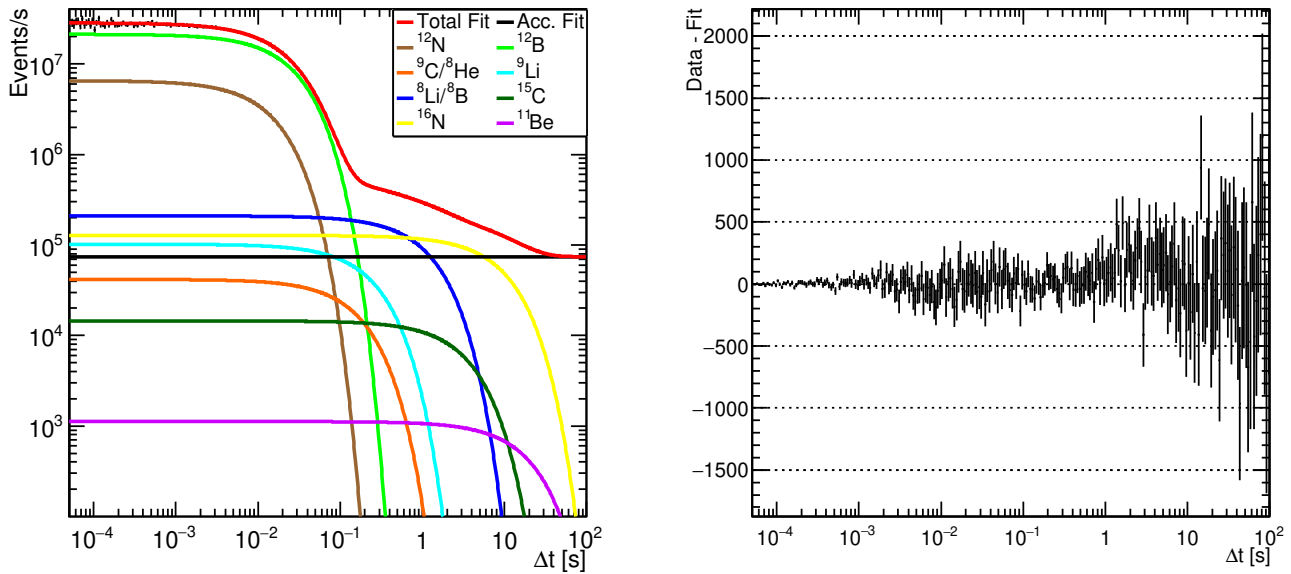}
    \caption{Left: Full $\Delta t$ distribution from 50~$\mu$s to 100~s. The solid red line corresponds to the entire fit, and the colored lines correspond to the individual isotope fits and background. Right: Fit residuals shown using unscaled data minus the fit value at the bin center multiplied by the with of the corresponding time bin.}
    \label{fig:fitandresidual}
\end{figure*}

In contrast to the the previous paper from~\cite{SKspall_zhang}, the rate distribution ($\Delta t$ distribution divided by time bin width) was made with logarithmic bins. Also, as mentioned earlier, the maximum $\Delta t$ considered for the fit was extended from 30~s to 100~s in order to account for long-lived isotopes and better constrain the distribution of accidentally coincident muon/spallation isotope candidate pairs. The result of this fit for all 10 isotopes is shown in Fig.~\ref{fig:fitandresidual}. The $\chi^2/NDF$ was 231.5/243 resulting in a p-value of 69\%. Figure~\ref{fig:fitandresidual} shows the fit to the entire time range as well as the residual of the fits. Overall, an excellent agreement between the data and the fit can be observed and the rates of the most abundant isotopes can be determined with percent-level precision.

\subsection{Spallation Efficiencies}
\label{sec:productionrates}
Obtaining isotope yields from the rates estimated above requires computing the efficiencies of the selection cuts described in~Sec.\ref{sec:preselyield}. These efficiencies were computed using both MC simulations and data. 

The efficiencies of the noise reduction, quality requirements, and energy cuts for the different spallation isotopes have been determined by simulating isotope decays using the SK detector simulation, based on GEANT 3.21~\cite{geant3}. In this simulation, we accounted for the time dependence of the detector properties using measurements of the PMT and water properties taken almost daily. The decay spectra are simulated using Geant~4.10.7~\cite{GEANT41,GEANT42} for most isotopes, taking into account both $\beta$ and $\gamma$ emission. Since the $^8$B decay was mismodeled in Geant4 we modeled its $\beta$ spectrum using tabulated values from~\cite{Winter:2004kf}. The final efficiencies are shown in Table~\ref{tab:spal_eff_simu}.

To determine the efficiency of the $l_t < 200$~cm cut introduced in Sec.~\ref{sec:preselyield}, we combine estimates from the data and from the spallation simulation described in Sec.~\ref{sec:simulation}. The simulation readily gives us efficiencies for individual isotopes paired with single through-going muons, that are shown in Table~\ref{tab:spal_eff_simu} but does not currently allow to study other muon categories.

\begin{table}[t]
    \centering
        \caption{Spallation sample efficiencies from simulation, for single through-going muons. The second column shows the isotope decay identification efficiencies for the FV, first reduction, and 6 MeV cuts, and the third column shows the efficiencies for the $lt \leq 200$~cm cut. The total $l_t$ efficiency has been obtained by averaging the efficiencies associated with the different isotopes, weighted by their predicted yields.}
    \begin{tabular}{ ccc }
    \toprule \multirow{2}{*}{Isotope}  & FV + 1st red & \multirow{2}{*}{$l_t \leq 200$~cm (\%)}\\
    & + E $\leq$ 6 MeV (\%) &  \\\hline
$\mathrm{^{12}N}$ & 69.0 & 94.9 \\ 
$\mathrm{^{12}B}$ & 54.3 & 89.9 \\ 
$\mathrm{^{8}He}$ & 23.9 & 94.6 \\ 
$\mathrm{^{9}C}$ & 67.4 & 94.0 \\ 
$\mathrm{^{9}Li}$ & 44.4 & 91.8 \\ 
$\mathrm{^{8}Li}$ & 51.8 & 91.4 \\ 
$\mathrm{^{8}B}$ & 48.1 & 92.9 \\ 
$\mathrm{^{15}C}$ & 37.1 & 87.7 \\ 
$\mathrm{^{16}N}$ & 54.8 & 88.9 \\ 
$\mathrm{^{11}Be}$ & 44.7 & 86.5 \\
\hline
Total & & 90.3 $\pm$ 2.8\\
   \bottomrule        
   \end{tabular}
    \label{tab:spal_eff_simu}
\end{table}

We complement this simulation study by extracting the $l_t$ cut efficiency from data for each muon type; we later take the total efficiency to be a weighted average using the observed relative contributions of each muon category.
We obtain the efficiencies for each muon type by fitting the $\Delta t$ distributions observed in 50~cm $l_t$ bins using the method described in Sec.~\ref{sec:yieldfit}. The 50~cm bin $\Delta t$ distribution were scanned to determine the largest $l_t$ where evidence for spallation could be found (the “endpoint”). The endpoints for each muon category, the $l_t$ efficiency and the spallation proportion is found in Table~\ref{tab:typeeff}. To combine the final $l_t$ efficiencies we then compute the relative contributions from each muon category to the spallation sample by considering all pairs within $l_t < 2000$~cm ---the maximal distance between spallation isotopes and their reconstructed parent muon track. The resulting total $l_t$ efficiency is calculated to be 73.6\%.

\begin{table}[t]
    \centering
        \caption{This table summarizes the proportion of total spallation production for the different muon categories and their respective efficiencies. 
    The proportion of spallation is measured as the ratio of the integral of the category fits at 2000 cm to the sum of all categories. }
    \begin{tabular}{>{\centering\arraybackslash}m{0.2\linewidth}>{\centering\arraybackslash}m{0.2\linewidth}>{\centering\arraybackslash}m{0.2\linewidth}>{\centering\arraybackslash}m{0.2\linewidth}}
         \toprule
         Category & Endpoint~[cm] & Efficiency~[\%] & Proportion~[\%]\\ \hline
         Single & 500 & 92.3 & 54.6 \\ \hline
         Multiple & 3000 & 49.3 & 43.5 \\ \hline
         Stopping & 400 & 93.4 & 1.9 \\ \hline
         Corner Clipping & N/A & N/A & $<10^{-3}$ \\ \hline 
         All & N/A & 73.6 & 100 \\ \bottomrule        
    \end{tabular}
    \label{tab:typeeff}
\end{table}

A second method identical to the previous results was implemented to validate: 
After applying a pre-cut on $\Delta t$ the efficiency is the ratio of the number of events within the $l_t$ cut value over all events passing the pre-cut. Eight different pre-cut values ranging from 10~ms to 30~s were chosen.
To estimate these efficiencies, a background sample was made using muons found after spallation candidates, as described in Sec.~\ref{sec:spaloglike}, and the $\Delta t$ distribution from this sample was subtracted from the spallation sample distribution. This procedure yields an $l_t$ efficiency of $74.2\pm0.5\%$. The total $l_t$ efficiency is then taken to be $74.0\pm0.7\%$ to account for the discrepancies between the two methods. The $l_t$ efficiency for single through-going muons obtained from data was compared to the average efficiency from the MC simulation and was found to be about $1\sigma$ away. 

\begin{table*}[t]
    \centering
        \caption{Observed and calculated spallation rates and yields using the $l_t$ cut. For the two sets of isotopes that could not be separated, the systematic error covers the range of yields corresponding to changing the relative fraction of an isotope from 0 to 1. The upper limit uses a 90\% confidence level~(C.L.) using the positive systematic error. Compared to the previous results, discovery for $^{15}\mbox{C}$ has been made, and much stronger constraint on the $^{11}\mbox{Be}$ measurement has been made. $^{11}\mbox{Be}$ was at a 1.5$\sigma$ excess. The $^{9}\mbox{C}$/$^{8}\mbox{He}$ and $^{9}\mbox{Li}$ fits for neutron clouds were 98.2\% anti-correlated. The total number of events for the two fit contributions is 65\% the corresponding $l_t^2$ contributions. The calculated relative fractions of $^{8}\mbox{Li}$ and $^{8}\mbox{B}$ in the  $^{8}\mbox{Li}+ {^{8}\mbox{B}}$ sample are 70.1\% and 29.9\% respectively. The calculated relative fractions of $^{9}\mbox{C}$ and $^{8}\mbox{He}$ in the  $^{9}\mbox{C} + {^{8}\mbox{He}}$ sample are 78.8\% and 21.2\% respectively. 
    }
    \begin{tabular}{c|cc|ccc|cc}
    \toprule   & \multicolumn{2}{c|}{\makecell{Yields \\ $[10^{-7}\mbox{cm}^2\mu^{-1}\mbox{g}^{-1}]$}} & \multicolumn{3}{c|}{Rates [$\mbox{kton}^{-1}\mbox{day}^{-1}$]} & \multirow{2}{*}{\makecell{Neutron data\\
    Fraction of \\ $l_t$ data}} & \multirow{2}{*}{\makecell{Calculated\\Isotope\\Fractions}}  \\
    Isotope  & Calculated & Observed & \makecell{Total Rate \\ ($l_t$ data)} & \makecell{Raw Rate \\ ($l_t$ data)} & \makecell{ Raw Rate \\ (neutron data)} & &\\\hline
$^{12}\mbox{N}$ & 0.92 &1.72 & $3.04\pm0.06\pm0.028$ &1.55  & 1.08 &  70\% & 2.3\% \\
$^{12}\mbox{B}$ & 8.6 & 12.9 & $22.86\pm0.11\pm0.21$  & 9.19 & 5.95 & 65\%  & 21.1\% \\
$^{9}\mbox{C}$/$^{8}\mbox{He}$ & 0.8 & \textless0.61 & \textless1.08 & 0.11 & 0.20 & 176\% & -- \\
$^{9}\mbox{Li}$ & 1.5 & 0.67 & $1.19\pm0.33\pm0.010$ &0.39  & 0.13 & 34\% & 3.7\% \\
$^{8}\mbox{Li}$/$^{8}\mbox{B}$ & 13.4 & 5.11 & $9.04\pm0.17^{+0.60}_{-1.1}$ & 3.69 & 2.59 & 70\% & 32.8\% \\
$^{15}\mbox{C}$ & 0.55 & 1.57 & $2.78\pm0.45\pm0.032$ & 0.76 & 0.37 & 49\% & 1.3\% \\
$^{16}\mbox{N}$ & 14.5 & 27.3 & $48.43\pm0.60\pm0.49$  & 19.64 & 12.01 & 61\% & 35.3\% \\
$^{11}\mbox{Be}$ & 0.61 & \textless1.05 & \textless1.9  & 0.33 & 0.19 & 56\% & -- \\
   \bottomrule        
   \end{tabular}
    \label{tab:isotopeyield}
\end{table*}
Finally, the isotope-dependence of the $l_t$ distribution was included to the measurement by scaling each isotope efficiency from MC by the weighted average of all isotopes. This factor was then used to scale the single through going measurement from data, and half of its effect was used to scale the multiple and stopping muon case. Since the multiple muon case was not performed in MC, this allowed for the difference to full or no isotope dependence to be covered in an isotope dependent systematic error. This error was relatively small for most isotopes, with only $^{15}\mbox{C}$ and $^{11}\mbox{Be}$ having an effect greater than the full $l_t$ systematic error.  
Here, using the simulation allows to refine the $l_t$ efficiency estimate performed in~\cite{SKspall_zhang}, where the isotope dependence was covered by a $\sim 4\%$ systematic uncertainties.

\subsection{Rate and Yields Calculation}
To calculate the total rates of the individual isotopes, the $l_t$ efficiency obtained from the data is combined with the efficiency associated with the noise, quality, and energy cuts. Raw rates for each isotope obtained from the fits in Sec~\ref{sec:yieldfit} are then corrected by these efficiencies to obtain the total production rates at SK-IV. For the isotopes that were paired together for the $\Delta t$ fit, the contributions of the isotopes were varied from 0 to 1 and used as a systematic error. These results are shown in Table~\ref{tab:isotopeyield}. 

The rates extracted from the data are compared to the FLUKA-based simulation described in Sec.~\ref{sec:simulation}. A simulated spallation sample of $1.362\times~10^8$ initial muons is generated in order to accumulate enough statistics for the low yield isotopes. The predicted rates are shown alongside the observed results in Table~\ref{tab:isotopeyield}.

Finally, the isotope yields are obtained by rescaling the total rates computed above for each isotope as follows:
\begin{equation}
    Y_i = \frac{R_i\cdot FV}{R_\mu\cdot\rho\cdot L_\mu}
\end{equation}
where $\rho$ is the density of water, $R_i$ is the total rate of the $i^{th}$ isotope at SK-IV, $R_\mu$ is the muon rate~(2.00~Hz), $L_\mu$ is the average length of reconstructed muon tracks, and $FV$ is the fiducial volume of the detector. Table~\ref{tab:isotopeyield} shows the final isotope yields for the data.

\subsection{Isotope study with neutron clouds}
\label{sec:neuteff}
In addition to updating the study performed in~\cite{SKspall_zhang}, we investigate the impact of neutron cloud cuts on isotope rates. Here, as stated in Sec.~\ref{sec:preselyield}, we consider a sample of spallation candidates paired with muons associated with at least three tagged neutrons. The fitting procedure described in Sec.~\ref{sec:yieldfit} is then performed for all isotopes, giving a $\chi^2$/dof of 252.6/243, which corresponds to a p-value of 0.323. Then, the final rates are scaled to account for the lower live time of the WIT trigger and allow a comparison with the $lt< 200$~cm sample. The scaled rates without efficiency corrections (also called raw rates) are shown in Table~\ref{tab:isotopeyield} for both the neutron cloud and the $l_t$ sample.

As shown in Table~\ref{tab:isotopeyield}, the fitted rates for the neutron cloud sample range between 33\% and 193\% of the rates found for the $l_t$ sample. The largest discrepancies between the two rates are seen for $^8$He/$^9$C and $^9$Li. For these subdominant isotopes, however, the precision of the fit is limited by statistics. Moreover, these isotopes have similar half-lives, which leads to degeneracies; the rates for the $^8$He/$^9$C and for $^9$Li are in fact found to be 98.3\% anti-correlated when considering the covariance of the fit parameters. For the most abundant isotopes, on the other hand, the ratio between the rates in the neutron cloud and $l_t$ sample remains around 60--70\%. The stability of this ratio highlights the correlation between the neutron cloud and $l_t$ cuts, as neutron cloud cuts also make use of the isotope distance to the muon track.

\section{conclusion}
\label{sec:conclusion}
In this paper, we presented new techniques to reduce spallation backgrounds for low energy analyses at SK, as well as the first realistic spallation simulation in the detector. We notably developed algorithms locating muon-induced hadronic showers, both by improving the reconstruction of the energy deposited along the muon track, and by identifying neutrons using a recently-implemented low energy trigger. New spallation cuts based on these algorithms allow to reduce the deadtime of the solar neutrino analysis by a factor of two, allowing a gain of the equivalent of one year of exposure at SK-IV. Moreover, the profiles of the neutron clouds produced in muon-induced hadronic showers are well reproduced by the spallation simulation, motivating its use to develop spallation reduction algorithms for future analyses. 

In addition to developing new spallation cuts, we computed the yields of the most abundant spallation isotopes at SK-IV, updating the study presented in~\cite{SKspall_zhang} with a 50\% increase in exposure. For the isotopes with the highest production rates, the yields can be determined with a precision of a few percent. Overall, the yields predicted by our spallation simulation lie within a factor of two of the observed values, well within the uncertainties associated with hadron production models. This study also demonstrated that identifying neutron captures associated with muons allows to build spallation-rich samples while keeping the relative fractions of the most abundant isotopes stable.

A central piece of the spallation studies described in this paper is the identification of neutrons produced in muon-induced showers. At SK-IV, the performance of our neutron tagging algorithm has been limited by the low livetime of the associated trigger, and the weakness of the neutron capture signal. At SK-Gd, however, the neutron capture visibility will be significantly increased due to gadolinium doping. Hence, algorithms based on neutron clouds will become a key component of the upcoming spallation reduction algorithms. In this context, the simulation presented in this paper will be instrumental in designing future analysis strategies. This paper thus demonstrates that, beyond neutrino-antineutrino discrimination, neutron tagging will impact significantly all low energy neutrino searches at SK.

\section*{Acknowledgments}
We gratefully acknowledge the cooperation of the Kamioka Mining and Smelting Company. The Super-Kamiokande experiment has been built and operated from funding by the Japanese Ministry of Education, Culture, Sports, Science and Technology, the U.S. Department of Energy, and the U.S. National Science Foundation. Some of us have been supported by funds from the National Research Foundation of Korea NRF-2009-0083526 (KNRC) funded by the Ministry of Science, ICT, and Future Planning and the Ministry of Education (2018R1D1A3B07050696, 2018R1D1A1B07049158), the Japan Society for the Promotion of Science, the National Natural Science Foundation of China under Grants No. 11620101004, the Spanish Ministry of Science, Universities and Innovation (grant PGC2018-099388-B-I00), the Natural Sciences and Engineering Research Council (NSERC) of Canada, the Scinet and Westgrid consortia of Compute Canada, the National Science Centre, Poland (2015/18/E/ST2/00758), the Science and Technology Facilities Council (STFC) and GridPPP, UK, the European Union’s Horizon 2020 Research and Innovation Programmeunder the Marie Sklodowska-Curie grant  agreement  no. 754496, H2020-MSCA-RISE-2018 JENNIFER2 grant agreement no.822070, and  H2020-MSCA-RISE-2019 SK2HK grant agreement no. 872549.
\appendix
\section{Simulation settings}
\label{appendix_sim}
In this appendix  the main settings chosen to build the FLUKA simulation are described in detail. 
FLUKA code fully integrates the most relevant physics models and libraries; it is not possible for the user to modify or adjust them according to their needs. Several default settings are available and must be chosen at the beginning of the simulation depending on the general physics problem the user is dealing with. In addition to this, FLUKA offers several options to customize the default settings enabling or disabling a certain type of processes or changing the treatment of specific type of interactions. For this work, FLUKA simulation was built with the default setting PRECISIO(n)~\cite{fluka_manual}. All the specifics that are particularly important for the scope of this paper are summarized below.

Low-energy neutron, which are defined to have less than 20~MeV energy, are transported down to thermal energies.

The absorption is fully analogue for low energy neutrons: in a fully analogue run, each interaction is simulated by sampling each exclusive reaction channel with its actual physical probability, this allows for event-by-event analysis. In general, this is not always the case, especially concerning low-energy neutron interactions.

Muon photonuclear interactions are activated with explicit generation of secondaries.

Several options are used to complement the default setting.

PHOTONUC option: photon and electron interactions with nuclei are activated at all energies.
 
MUPHOTON option: controls the full simulation of muon nuclear interactions at all energies and the production of secondary hadrons.

EVAPORAT(ion) and COALESCE(nce) options: these two are set to give a more detailed treatment of nuclear de-excitations. Despite the related large CPU penalty, it is fundamental to activate these options when isotope production want to be studied. EVAPORAT enables the production of heavy nuclear fragments ($A>1$) while COALESCE sets the emission of energetic light-fragments. 

IONSPLIT option: used for activating ion splitting into nucleons.

IONTRANS option: full transport of all light and heavy ions and activation of nuclear interactions.

RADDECAY option: activate radioactive decay calculations.

Settings are specified in the main input file: FLUKA, unlike other Monte Carlo particle transport codes, is built to get the basic running conditions from a single standard code. However, due to the complexity of spallation mechanism, standard options do not satisfy to retrieve the problem-specific informations we need to score: customized input and output routines (SOURCE and MGDRAW) are required to be written in order to incorporate non-standard primary particle distributions, the ones calculated with MUSIC simulation, and to extract event-by-event informations for the shower reconstruction. In particular, only primary muons inducing the production of at least one hadron or of an isotope are selected and recorded; the rest are not interesting for this study and are discarded to save computational time.
\section{Spallation Likelihood Fit Parameters}
\label{appendix_fitparam}

In the following section of the appendix, all of the fit parameters for the different components of the $\log_{10}\mathcal{L}$ will be included. The
fit to the random coincidence sample is referred to as ``random'', while ``spallation'' means the fit to the random coincidence-subtracted spallation sample. For contributing variables with only one spallation and random coincidence function ($\Delta l_{\mbox{\tiny LONG}}$, $\Delta t$, and $Q_{res}$) there is no normalization factor while for $lt^2$ a normalization factor is needed to due to the multiple bins for the function. The normalization factors are not listed.

\subsection{$\Delta t$}
There is only a spallation function as follows as the constant fit for random coincidences is dropped:
\begin{equation}
    \mbox{\it PDF}_{sig}(\Delta t) = \sum_i^7 A_i e^{-\Delta t/\tau_i}
\end{equation}
where $\tau_i$ is the decay constant for the isotope and $A_i$ is the fitted amplitude. The fit parameters are listed in Table~\ref{tab:fitdt}.
\begin{table}[h]
    \centering
        \caption{$\Delta t$ fit parameters}
    \begin{tabular}{cccc}
        \toprule
         Exponential $i$ & $A_i$ & $\tau_{i}$~[s] & Isotopes \\ \hline
         1 & 23500 & 0.0159 & $^{12}$N, $^{13}$O, $^{11}$Li\\
         2 & 83250 & 0.02943 & $^{12}$B, $^{13}$B, $^{14}$B\\
         3 & 234.7 & 0.2568 & $^9$Li, $^9$C \\
         4 & 869.2 & 1.212 & $^8$Li, $^8$B, $^{16}$C\\
         5 & 93.37 & 3.533 & $^{15}$C \\
         6 & 468.6 & 10.29 & $^{16}$N \\
         7 & 5.400 & 19.91 & $^{11}$Be \\ \bottomrule        
    \end{tabular}
    \label{tab:fitdt}
\end{table}

\subsection{Transverse Distance}

\begin{table*}[hbt]
\centering
\caption{Spallation and Random PDF fit parameters for $l_t^2$.}
\begin{tabular}{cc|cccccc|cccc}
\toprule
& & \multicolumn{6}{c|}{Spallation} & \multicolumn{4}{c}{Random} \\
\hline
Time  & resQ    & $c_1$ & $p_1$ & $c_2$ & $p_2$   & $c_3$ & $p_3$ &
$l_{t0}^2$ & $p_0$ & $p_1$      & $p_2$ \\
~[s] & [1000~p.e.] &  & [m$^{-2}$] &   &  [m$^{-2}$] &  & [m$^{-2}$] &
[m$^2$]  &       & [(100~m)$^{-2}$] & [(100~m)$^{-4}$] \\
\hline
      & $<0$    & 8.023 & 5.506 & 7.678 & 2.082   & 5.180 & 0.4476  &
      0  & 8.035  & 35.09 & 140.6  \\
      & 0--25    & 8.929 & 4.170 & 8.145 & 1.511   & 5.329 & 0.3031  &
      64 & 11.19  & 16.27 & $-75.68$ \\
      & 25--50   & 8.354 & 4.860 & 7.278 & 1.348   & 4.696 & 0.2588  &
      0  & 2.638  & 30.20 &  11.59 \\
0--0.1 & 50--100  & 8.395 & 2.820 & 6.036 & 0.5158  & 1.774 & 0.03065 &
      0  & 2.005  & 35.58 & 114.2  \\
      & 100--500 & 8.770 & 2.248 & 6.129 & 0.3654  & 2.712 & 0.01120 &
      0  & 1.954  & 35.02 &  71.72 \\
      & 500--1000& 6.869 & 1.906 & 4.353 & 0.2588  & 1.713 & 0.00967 &
      0  & 0.1952 & 36.25 &   0    \\
      & $>1000$ & 5.772 & 2.542 & 4.728 & 0.3412  & 1.900 & 0.00755 &
      0  & 0.2675 & 61.74 &   0    \\
\hline
      & $<0$    & 8.065 & 4.446 & 5.021 & 0.4160  & 7.063 & 1.773   &
      0  & 273.2 & 26.02 & $-47.17$  \\
      & 0--25    & 8.758 & 3.067 & 6.683 & 0.6694  & $-16.92$ & 0.02444 &
      16 & 365.6 & 10.89 &$-107.8$   \\
      & 25--50   & 7.543 & 4.283 & 7.361 & 1.656   & 4.470 & 0.2228  &
      9  & 75.61 & 19.93 &$-115.7$   \\
0.1--3 & 50--100  & 8.087 & 3.467 & 6.093 & 0.6448  & 2.574 & 0.06516 &
      4  & 60.11 & 23.04 &$-146.0$   \\
      & 100--500 & 8.377 & 2.684 & 6.633 & 0.5569  & 2.579 & 0.01071 &
      0  & 64.97 & 28.45 &$-165.6$   \\
      & 500--1000& 6.533 & 1.847 & 4.544 & 0.3578  & 1.464 & 0.01112 &
      0  & 7.004 & 45.45 &  $-1.876$ \\
      & $>1000$ & 5.587 & 1.997 & 4.139 & 0.2551  & 2.013 & 0.01305 &
      0  & 3.265 & 53.02 &$-199.6$   \\
\hline
      & $<0$    & 8.742 & 2.810 & 6.099 & 0.4985  &$-100117$ & 349594  &
      0  & 2557  & 26.18 & $-39.10$  \\
      & 0--25    & 2.611 &0.00484& 7.220 & 0.6481  & 8.999 & 3.075   &
      16 & 3377  & 10.11 & $-115.0$   \\
      & 25--50   & 8.080 & 3.452 & 7.078 & 0.8928  & 3.131 & 0.03730 &
      0  & 719.2 & 19.35 &$-104.9$   \\
3--30  & 50--100  & 8.071 & 2.845 & 6.805 & 0.8500  & 3.284 & 0.03119 &
      0  & 571.7 & 23.95 &$-103.2$   \\
      & 100--500 & 8.656 & 2.048 & 6.302 & 0.3115  & 2.164 & 0.00498 &
      0  & 630.7 & 29.57 & $-120.8$   \\
      & 500--1000& 6.782 & 1.465 & 1.451 & 0.00314 & 4.601 & 0.1978  &
      0  & 62.35 & 35.25 & $-149.7$   \\
      & $>1000$ & 5.885 & 1.994 & 2.999 & 0.01909 & 4.521 & 0.03328 &
      64 & 26.68 & 56.73 & 161.6   \\
\bottomrule        
\end{tabular}
\label{tab:fitlt2}
\end{table*}

The fit for $l_t^2$ is carried out over 7 residual charge bins and 3 time ranges, corresponding to 21 different fits. The equations for the spallation and random coincidence fits are as follows:
\begin{gather}
    \mbox{\it PDF}_{spa,l_t}(l_t^2) = \sum_{i=1}^3e^{c_i - p_i\cdot l_t^2}\\
    \mbox{\it PDF}_{rnd,l_t}(l_t^2) = 
    \begin{cases}
       p_0 & l_t^2 \leq l_{t0}^2 \\ 
       p_0e^{-p_1(l_t^2 - l_{t0}^2) + p_2(l_t^2-l_{t0}^2)^2}          & l_t^2 > l_{t0}^2 
    \end{cases}
\end{gather}
where $c_i$ and $p_i$ are the fit parameters (see Table~\ref{tab:fitlt2}). Due to the finite size of the detector, at large $l_t$ the allowed region is no longer cylindrical, so a ``piecewise'' function was defined for the random PDFs, with $l_{t0}^2$ being the point where the function changes. The parameters are listed in Table~\ref{tab:fitlt2}.

\subsection{Residual Charge}

\begin{table}[hbt]
\centering
\caption{PDF fit parameters for residual charge.}
\begin{tabular}{cccc}
\toprule
PDF & Exponential $i$ & $c_i$ & $p_i$ [(1000~p.e.)$^{-1}$] \\
\hline

           & 1 & $-10.134$ & $6.480$    \\
           & 2 & $-0.745$ & $0.08697$ \\
Spallation & 3 &  $1.424$ & $0.01459$ \\
           & 4 & $-2.722$ & $0.00396$ \\
           & 5 & $-4.801$ & $0.00097$ \\
\hline
           & 1 &  $4.507$ & $0.1289$   \\
           & 2 &  $1.990$ & $0.02945$  \\
Random     & 3 &  $0.328$ & $0.01099$  \\
           & 4 & $-2.257$ & $0.003443$ \\
           & 5 & $-4.977$ & $0.001091$ \\
\bottomrule        
\end{tabular}
\label{tab:fitresq}
\end{table}

Both signal and background PDFs in the positive $Q_{res}$ region are a sum of exponential functions:
\begin{equation}
\mbox{\it PDF}(Q_{res}) =\sum^5_{i=1} e^{c_i - p_i\cdot Q_{res}}, \ \ Q_{res} > 0
\end{equation}
where $c_i$ and $p_i$ are the fit parameters for the exponential functions.

For negative $Q_{res}$, no analytical form was assumed and a linear interpolation of the sample bins was used between the four points listed in Table~\ref{tab:fitnegresq}.

\begin{table}[hbt]
\centering
\caption{PDF points for negative residual charge. Residual charges below $-100,000$ are not considered.}
\begin{tabular}{cccc}
\toprule
PDF & Point & Residual charge~[1000~p.e.] & PDF value\\
\hline
Spallation & 1 &  $-100$   & $10^{-6}$ \\
           & 2 &   $-65$   & $0.000713$ \\
           & 3 &   $-22.5$ & $0.02107$  \\
           & 4 &   $-7.5$ & $0.34053$   \\
\hline
Random     & 1 &  $-100$   & $10^{-6}$\\
           & 2 &   $-65$   & $0.07034$ \\
           & 3 &   $-22.5$ & $0.609$   \\
           & 4 &   $-7.5$  & $23.29$  \\
\bottomrule        
\end{tabular}
\label{tab:fitnegresq}
\end{table}

\subsection{Longitudinal Distance}

Here, the PDFs are modeled as triple (double) Gaussians for spallation (random coincidence):
\begin{equation}
\mbox{\it PDF}(\Delta l_{\mbox{\tiny long}}) = \sum^3_{i=1} A_ie^{\frac{-(\Delta l_{\mbox{\tiny long}} - x_i)^2}{2\sigma_i^2}}
\end{equation}
where $A_i$, $x$, and $\sigma_i$ are the amplitude, offset and width parameters for each Gaussian. For Table~\ref{tab:fitsigln} lists the parameters.

\begin{table}[h]
\centering
\caption{PDF fit parameters for $\Delta l_{\mbox{\tiny LONG}}$.}
\begin{tabular}{ccccc}
\toprule
PDF & Gaussian & amplitude & offset~[m] & width~[m] \\
\hline

           & 1 & $989.8$ & $-3.090$ & $0.9217$ \\
Spallation & 2 & $1185$  &  $26.32$ & $1.852$  \\
           & 3 & $254.6$ & $-58.87$ & $13.49$  \\
\hline
\multirow{2}{*}{Random} &
             1 & $150.0$ & $-1.160$ & $13.41$  \\
           & 2 & $20.18$ & $-25.16$ & $3.976$ \\
\bottomrule        
\end{tabular}
\label{tab:fitsigln}
\end{table}

\bibliography{references}
\end{document}